\documentclass[11pt,a4paper]{article}
\usepackage{epsfig}
\usepackage[T1]{fontenc}    
\usepackage{graphics}
\usepackage{graphicx}
\usepackage{pstricks,pst-coil,pst-fill,pst-plot}
\usepackage[fleqn]{amsmath}    
\usepackage{amssymb}    
\usepackage{amsfonts}   
\usepackage{verbatim}   
\usepackage{mathrsfs}   
\usepackage{dsfont}
\usepackage{euscript}
\usepackage{yfonts}
\usepackage{txfonts}
\usepackage{marvosym}
\usepackage{vmargin}        

\setmarginsrb{1.8cm}{2cm}{1.8cm}{2cm}{1cm}{1cm}{1cm}{1.6cm}
 \makeatletter
 \@addtoreset{equation}{section}
 \makeatother

\let\tend=\rightarrow

\newtheorem{theorem}{Theorem}[section]
\newtheorem{prop}{Proposition}[section]

\newtheorem{rem}{Remark}[section]

\def\Proof{\medskip\noindent {\it Proof --- \ }}

\def\qed{\hfill\rule{2mm}{2mm}}

\newcommand\beq{\begin{equation}}
\newcommand\enq{\end{equation}}
\newcommand\bem{\begin{multline}}
\newcommand\enm{\end{multline}}

\def\beqa{\begin{eqnarray}}
\def\eeqa{\end{eqnarray}}
\def\ba{\begin{array}}
\def\ea{\end{array}}

\def\det{\operatorname{det}}

\newcommand{\f}[2]{{\ensuremath{%
    \mathchoice%
    {\dfrac{#1}{#2}}
    {\dfrac{#1}{#2}}
    {\frac{#1}{#2}}
    {\frac{#1}{#2}}
}}}
\newcommand{\tf}[2]{\ensuremath{#1/#2}}
\newcommand{\pa}[1]{\ensuremath{\left(#1\right)}}

\newcommand{\pac}[1]{\ensuremath{\left[#1\right]}}

\newcommand{\pab}[2]{\ensuremath{\pa{\ba{c} #1 \\ #2 \ea }}}

\def\be{\beta}
\def\ga{\gamma}
\def\Ga{\Gamma}
\def\de{\delta}

\def\De{\Delta}

\def\la{\lambda}

\def\sg{\sigma}

\def\Sg{\Sigma}

\def\Ups{\Upsilon}

\def\Om{\Omega}

\newcommand{\mc}[1]{\ensuremath{\mathcal{#1}}}

\newcommand{\bs}[1]{\ensuremath{\boldsymbol{#1}}}

\newcommand{\ov}[1]{\ensuremath{\overline{#1}}}

\newcommand{\Int}[2]{\ensuremath{\int\limits_{#1}^{#2}}}
\newcommand{\Oint}[2]{\ensuremath{\oint\limits_{#1}^{#2}}}

\newcommand{\sul}[2]{\ensuremath{\sum\limits_{#1}^{#2}}}

\newcommand{\R}{\ensuremath{\mathbb{R}}}
\newcommand{\Cx}{\ensuremath{\mathbb{C}}}

\newcommand{\Dp}[1]{\ensuremath{\partial_{#1}}}

\newcommand{\limit}[2]{\ensuremath{\underset{#1 \tend #2}{\longrightarrow} }}

\newcommand{\ex}[1]{\ensuremath{\e{e}^{#1}}}

\newcommand{\ddet}[2]{\ensuremath{\det_{#1}\pac{#2}}}

\newcommand{\abs}[1]{\ensuremath{\left| #1 \right|}}

\newcommand{\dd}{\mathrm{d}}
\newcommand{\e}[1]{\ensuremath{\mathrm{#1}}}

\newcommand{\intff}[2]{\ensuremath{\left [ \, #1 \,; #2 \, \right ] }}
\newcommand{\intfo}[2]{\ensuremath{\left [ \, #1 \,; #2 \, \right ) }}
\newcommand{\intof}[2]{\ensuremath{\left ( \, #1 \,; #2 \, \right ] }}
\newcommand{\intoo}[2]{\ensuremath{\left ( \, #1 \,; #2 \, \right ) }}

\newcommand{\C}{{\mathbb C}}

\newcommand{\ep}{\varepsilon}

\newcommand{\di}{\displaystyle}

\begin{document}

\begin{flushright}

\end{flushright}
\par \vskip .1in \noindent

\vspace{14pt}

\begin{center}
\begin{LARGE}
{\bf Riemann-Hilbert approach to a generalised sine kernel.}
\end{LARGE}

\vspace{30pt}

\begin{large}

{\bf R.~Gharakhloo}\footnote[1]{Department of Mathematical Sciences,
 Indiana University-Purdue University,
 402 N. Bla\-ckford St.,
 Indianapolis, IN 46202-3267,
 USA, rgharakh@iu.edu}

{\bf A.~R.~Its}\footnote[1]{Department of Mathematical Sciences,
 Indiana University-Purdue University,
 402 N. Bla\-ckford St.,
 Indianapolis, IN 46202-3267,
 USA, aits@iupui.edu}, \footnote[2]{St. Petersburg State University, Universitetskaya emb. 7/9, 199034, St. Petersburg,
Russia.}
\par

{\bf K.~K.~Kozlowski}\footnote[3]{Univ Lyon, ENS de Lyon, Univ Claude Bernard Lyon 1, CNRS, Laboratoire de Physique, F-69342 Lyon, France, 
karol.kozlowski@ens-lyon.fr}. 
\par

\end{large}

\vspace{40pt}

\centerline{\bf Abstract} \vspace{1cm}
\parbox{12cm}{\small 
We derive the large distance asymptotics of the Fredholm determinant of the so-called generalised sine kernel 
at the critical point. This kernel corresponds to a generalisation of the pure sine kernel arising in the theory of random matrices and has 
potential applications to the analysis of the large-distance asymptotic behaviour of the so-called emptiness formation 
probability for various quantum integrable models away from their free fermion point.}

\end{center}

\vspace{40pt}

\section*{Introduction}

\vspace{10pt}

Correlation functions in quantum integrable systems at their free fermion point are represented by Fredholm determinants 
(or minors theoreof) of integrable integral operators \cite{ItsIzerginKorepinSlavnovDifferentialeqnsforCorrelationfunctions}. In this setting, the kernels of these operators depend, in an 
oscillatory way, on the spacial (and/or temporal) separation between the various operators whose correlation function is being computed. 
Various physical reasons such as the testing of universality in the large-distance $x$ (and/or temporal $t$)
regime motivate the analysis of the large $x$ (and/or $t$) behaviour of the underlying Fredholm determinants. 
The setting of a Riemann--Hilbert problem \cite{ItsIzerginKorepinSlavnovDifferentialeqnsforCorrelationfunctions}
associated with these operators alongs with the development of the Deift-Zhou non-linear steepest descent method
\cite{DeiftZhouSteepestDescentForOscillatoryRHP} led to a complete understanding of the asymptotical regimes of correlation functions at the 
free fermion point. The situation changes drastically in the case of quantum integrable systems that are away from their 
free fermion point. Indeed, then, the representation for the correlation functions become much more
complicated and take the form of  so-called multidimensional Fredholm series 
\cite{KozKitMailSlaTerXXZsgZsgZAsymptotics,KozReducedDensityMatrixAsymptNLSE,KozTerNatteSeriesNLSECurrentCurrent}.
These can be seen as certain very specific deformations of the Fredholm series for a Fredholm determinant which do preserve some of the features of the integrands arising in the Fredholm series. 
Although, the general theory of these entities has not yet been fully established, there has already emerged a certain number of efficient
methods allowing one to study, under certain hypotheses, the asymptotic behaviour of certain examples of such series \cite{KozKitMailSlaTerXXZsgZsgZAsymptotics,KozReducedDensityMatrixAsymptNLSE,KozTerNatteSeriesNLSECurrentCurrent}. 
In particular, the method  of multidimensional deformations of a Natte series allows one to prove the asymptotic expansion of the correlator
\textit{under} the assumption of the convergence of the multidimensional Natte series of interest. 
The proof of the convergence of such series is a hard open problem that is, however, disconnected from the 
asymptotic analysis \textit{per se}. The scheme of this method goes as follows. 
One starts by identyfing the integrable integral operator that is underlying to the multidimensional Fredholm series of interest.  
The next step consists in studying, with the help of Riemann--Hilbert problem-based methods, the asymptotic behaviour of the Fredholm determinant associated with this operator. Having obtained these,
one should then raise them to the level of the multidimensional Fredholm series by using the techniques developed in 
\cite{KozReducedDensityMatrixAsymptNLSE,KozTerNatteSeriesNLSECurrentCurrent}. 
Of a particular interest is a specific correlation function, called the "emptiness formation probability", which arises  
 in quantum integrable models and has a similar interpretation to the gap probability arising in matrix models. 
The multidimensional Fredholm series for this correlator has been obtained in \cite{KozKitMailSlaTerXXZsgZsgZAsymptotics}.
The integrable kernel coming up  in the aforestated scheme of asymptotic analysis corresponds to the generalised sine kernel at the critical point. 
It is this very reason that triggered our interest 
in this kernel. 

We shall now be more precise. By generalised sine kernel at the critical point, we mean the integral operator 
$I+V$ acting on $L^{2}(\intff{a}{b})$, $a,b \in \R$, $a<b$, with the integral kernel 
\beq
V(\la,\mu) \; = \; - \f{  e(\la)e^{-1}(\mu) - e(\mu)e^{-1}(\la) }{ 2i\pi (\la- \mu) }  
\qquad \e{where} \quad  e(\la)  \; = \;  \ex{ \f{im}{2} p(\la) \, + \, \f{g(\la)}{2} }  \;. 
\label{definition noyau GSK}
\enq
In the following, we shall assume that 
\begin{itemize}
\item $p^{\prime}>0$ on $\intff{a}{b}$ ; 
\item $p$ and $g$ are holomorphic on some open neighbourhood $U$ of $\intff{a}{b}$. 
\end{itemize}
Note that, in the large-$m$ regime, one may absorb $g$ in the definition of $p$ : $p(\la) \hookrightarrow p_g(\la) = p(\la) - \tf{g(\la)}{m}$. 
However, treating $g$ on separate grounds leads to a different representation for the asymptotic expansion of the solution to the 
associated Riemann--Hilbert problem and, hence, of the determinant. It is this alternative form of the asymptotic expansion 
that would be more useful from the point of view of the potential applications of our analysis to large-distance asymptotics of the emptiness formation probability. 

As it has already been  mentioned, our interest to the integrable integral operators with kernel \eqref{definition noyau GSK} is primary
motivated by the asymptotic problems related to the emptiness formation probability in the non-free fermion exactly solvable quantum models.
Hence our motivation to devote the last section  of the paper to one special  variant of the sine kernel  which appears in the description of the emptiness formation probability in the 
XXZ spin-1/2 Heisenberg chain. This generalisation  can be, of course, treated as a particular case of the kernel 
\eqref{definition noyau GSK}; however, it is not directly in this class. Indeed, in that case, the integral operator acts on functions supported on an arc instead of an interval
so that some modifications of the analysis are needed. This particular instance of the generalised sine kernel is also a deformation of the
specific integrable kernel of sine-type considered in \cite{dik}. Moreover, this deformation does not affect much the analysis of \cite{dik},
so that, after deriving the relevant differential identity for the determinant in question, we can just use, with  a proper adjustment,
the results of \cite{dik} and produce the asymptotics of this specific determinant. We hope that the results of this last section can
be further used for the final evaluation of the asymptotics of the XXZ emptiness formation probability. For the reader's convenience, at the beginning of the last section, we discuss  in more details the 
definition of the emptiness formation probability in the XXZ chain.

We start the paper by carrying out an asymptotic in $m$ resolution of the Riemann--Hilbert problem associated with the kernel 
\eqref{definition noyau GSK} in Section \ref{Section Resolution Asympt RHP}. Then, in Section \ref{Section asymptotic evaluation of the determinant}
we build on the previous analysis so as to estimate the Fredholm determinant $\det\big[I+V\big]$ asymptotically in $m$. 
Finally, in Section \ref{efp}, we discuss the asymptotics of the Fredholm determinant directly connected with the XXZ chain.

\section{Asymptotic resolution of the Riemann--Hilbert problem}\label{genas}
\label{Section Resolution Asympt RHP}

\subsection{The kernel and initial Riemann--Hilbert problem}
\label{SousSection Initial RHP}

The kernel \eqref{definition noyau GSK} is of integrable type in that it admits the representation:
\beq\label{intkernel}
V(\la,\mu) =  \f{  \big( \bs{E}_{L}(\la ) , \bs{E}_{R}(\mu) \big)  }{ \la-\mu}\,,   
\enq
with 
\beq
 \bs{E}_{L}^{\bs{T}}(\la ) \; = \;     \big( - e^{-1}\!(\la) \; \;   e(\la) \big)\,,  \qquad \e{and} \qquad 
  \bs{E}_{R}(\la) \; = \;  \f{-1}{2i\pi}  \pa{\ba{c} e(\la) \\ e^{-1}\!(\la) \ea}  \;. 
\enq
Above, $\bs{T}$ in the exponent refers to the transposition.

\noindent This kernel is associated with the Riemann--Hilbert problem \cite{ItsIzerginKorepinSlavnovDifferentialeqnsforCorrelationfunctions}
for a $2\times 2$ matrix $\chi(\la)$
\begin{itemize}
\item $\chi \in \mc{O}(\Cx\setminus \intff{ a }{ b })$ and has continuous boundary values on 
$\intoo{ a }{ b }$;
\item $\chi(\lambda)=\ln\abs{\lambda-a} \cdot \ln\abs{\lambda-b} \cdot \e{O}\pa{ \ba{cc} 1 & 1 \\ 1 & 1 \ea}$ 
\; \;  as $\lambda \tend a$ \; \;  or \; \;  $\lambda \tend b$; 
\item $\chi(\lambda) = I_{2} \; + \; \e{O}\big(\lambda^{-1}\big)$ when $\lambda \tend \infty$;
\item $\chi_-(\la) = \chi_+(\la) \cdot  G_{\chi}(\la)$ for $\la \in \intoo{a}{b}$, \; \;  where 
\begin{equation}\label{genchiRH}
G_{\chi}(\la) = I_2 + 2i\pi  \bs{E}_{R}(\la) \cdot  \bs{E}_{L}^{\bs{T}}(\la)
\end{equation}
\end{itemize}
Here, $\mc{O}(U)$, $U$ open in $\Cx$, stands for the ring of holomorphic functions on $U$. $I_2$ is the identity matrix. Also, we 
should explain that relations of the type $M(z) = \e{O}(N(z))$ for two matrix functions $M,N$ should be understood entrywise, 
\textit{i.e.} $M_{jk}(z)= \e{O}\big( N_{jk}(z) \big)$. Further, given a function $f$ defined on $\Cx\setminus \ga$, with $\gamma$
an oriented curve in $\Cx$, we denote by $f_{\pm}(s)$  the boundary values of $f(z)$ on $\Ga$ when the argument $z$ 
 approaches the point $s \in \Ga$ non-tangentially and from the right ($+$) or the left ($-$) side of the curve. Again, if one deals with 
matrix function, then this relation has to be understood entrywise. 

The unique solution to the above Riemann--Hilbert problem takes the form
\beq
\chi(\la) = I_2 - \Int{a}{b}  \f{  \bs{F}_{R}(\mu) \cdot  \bs{E}_{L}^{\bs{T}}(\mu) }{ \mu- \la }  \dd \mu\,,    \qquad \e{and} \qquad
\chi^{-1}(\la) = I_2 + \Int{a}{b}  \f{ \bs{E}_{R}(\mu) \cdot  \bs{F}_{L}^{\bs{T}}(\mu) }{ \mu - \la }  \dd \mu\,, 
\label{formules reconstruction chi chi-1 en terms F R et FL}
\enq
where $\bs{F}_{R}(\la) $ and $\bs{F}_{L}(\la) $ correspond to the solutions to the below linear integral equations
\beq
\bs{F}_{R}(\la) \; + \; \Int{a}{b} V(\mu,\la) \bs{F}_{R}(\mu)d\mu \; =\;   \bs{E}_{R}(\la)\,,
\qquad \e{and} \qquad 
\bs{F}_{L}(\la)  \; + \; \Int{a}{b} V(\la,\mu)\bs{F}_{L}(\mu)d\mu  \; =  \;  \bs{E}_{L}(\la) \;. 
\enq

The jump matrix for $\chi$ has a 0 in its lower diagonal:
\beq
G_{\chi}(\la) = \pa{\ba{cc} 2  & -e^{2}(\la)  \\
							e^{-2}(\la) & 0 \ea } \;. 
\enq

\subsection{The first transformation: $h$-function}\label{sec1.2}

The function
\beq
h(\la) = q(\la) \Int{a}{b} \f{ -ip(s) }{ q_+(s)(s-\la) }  \f{ \dd s }{ 2i\pi } \qquad \e{with} \qquad 
q(\la) = (\la-a)^{\f{1}{2}} (\la-b)^{\f{1}{2}}\;
\enq
solves the Riemann--Hilbert problem
\beq
h\in \mc{O}(\Cx \setminus \intff{a}{b} ) \qquad ,  \qquad h_{+}(\la) \; + \;  h_-(\la) = -ip(\la) \;
\qquad \e{for} \qquad \la \in  \intoo{a}{b}
\enq
and $h$ is bounded at infinity. Moreover
\beq
h(\la) \limit{ \la }{ \infty }  h_{\infty}  = \f{1}{2\pi} \Int{ a }{ b } \f{p(s)}{q_+(s)} \dd s \;. 
\enq

Let $\Ga$ be a counterclockwise loop around $\intff{a}{b}$ in $U$. If $\la$ belongs to $U$ and is located 
outside of the loop $\Ga$, then it is readily seen that $h$ admits the representation 
\beq
h(\la) =   q(\la) \Oint{\Ga}{} \f{p(s)}{  s-\la  } \f{\dd s }{  4 \pi q(s) }  \;. 
\label{ecriture rep int h pour la exterieur ctr}
\enq
Furthermore, if $\la$ is located inside of 
$\Ga$, then $f(\la)$ admits the alternative representation
\beq
h(\la) =   q(\la) \Oint{\Ga}{} \f{p(s)-p(\la)}{  s-\la  } \f{\dd s }{  4 \pi q(s) } 
 \; - \; i \f{ p(\la) }{ 2 } \;. 
\label{ecriture rep int h pour la int ctr}
\enq
Here, the regularising term could have been added since the corresponding integral is zero (the residue at infinity vanishes). 
It is then easy to see that,  for $\la \in \intoo{a}{b}$,
\beq
h_-(\la) \, - \, h_+(\la)   \; = \; (\la-a)^{\f{1}{2}} \cdot (b-\la)^{\f{1}{2}}  \Int{a}{b}  \f{p(\la)-p(s)}{\pi(\la-s)} 
\f{\dd s}{ (s - a)^{\f{1}{2}} \cdot (b - s)^{\f{1}{2}}  }  >0\;,
\enq
where the positivity follows from 
\beq
\Int{a}{b}  \f{p(\la)-p(s)}{\pi(\la-s)} 
\f{\dd s}{ (s - a)^{\f{1}{2}} \cdot (b - s)^{\f{1}{2}}  } > \inf_{s\in \intff{a}{b}}[ p^\prime(s) ]  \; \cdot \;
 \Int{a}{b} \f{\dd s}{ \pi (s - a)^{\f{1}{2}} \cdot (b - s)^{\f{1}{2}}  }  \; >0 \;. 
\enq

We then set 
\beq
\Xi(\la) = \ex{m h_{\infty} \sg_3} \chi(\la) \ex{ - m h(\la) \sg_3 } \;. 
\enq
It is readily seen, that
\begin{itemize}
\item $\Xi \in \mc{O}\big(\Cx\setminus \intff{a}{b} \big)$ and has continuous boundary values
on $\intoo{a}{b}$;
\item $\Xi(\la)=\ln\abs{\la-a} \cdot \ln\abs{\la-b} \cdot \e{O}\pa{ \ba{cc} 1 & 1 \\ 1 & 1 \ea}$ 
\; \; as $\la \tend a$ \;\;  or \;\;  $\la \tend b$; 
\item $\Xi(\la) = I_{2} \; + \; \e{O}\big(\la^{-1}\big)$ when $\la \tend \infty$;
\item $\Xi_-(\la) = \Xi_+(\la)\cdot G_{\Xi}(\la)$ for $\la \in \intoo{a}{b}$ where 
\beq
\qquad G_{\Xi}(\la) = \pa{ \ba{cc} 2 \ex{-m [h_-(\la) - h_+(\la)] }  &  - \ex{g(\la)} \\ 
						\ex{-g(\la)}  &  0 \ea }  \;. 
\label{XiRH}
\enq
\end{itemize}

\subsection{The parametrix on $\intff{a}{b}$} \label{glpar}

The scheme for building the global parametrix  on $\intff{a}{b}$ is standard
\cite{DeiftItsZhouSineKernelOnUnionOfIntervals,KuilajaarsMVVUniformAsymptoticsForModifiedJacobiOrthogonalPolynomials}. We set 
\beq\label{DDinfty}
D(\la) = \exp \Bigg\{ q(\la)\Int{a}{b} \f{ -g(s) }{ q_+(s)} \f{\dd s}{2i\pi(s-\la)}  \Bigg\}
\qquad \e{and} \qquad 
D_{\infty} = \exp \Bigg\{ \Int{a}{b} \f{ g(s) }{ 2i\pi q_+(s)} \dd s  \Bigg\} \;. 
\enq
This can be alternatively recast as
\beq
D(\la) = \exp \Bigg\{ q(\la)\Oint{ \Ga }{} \f{ g(s) }{ q(s) (s-\la)  } \f{\dd s}{4i\pi}  
		\; - \; \f{ g(\la) }{ 2 } \Bigg\}  \;. 
\enq
In which $\Ga$ refers to a loop around $\intff{a}{b}$ in $U$ that, furthermore, encircles $\la$.
Hence, one has
\beq
D_+(\la) D_-(\la) = \ex{-g(\la)} \;, \qquad \la \in \intoo{a}{b} \;. 
\enq
Agreeing upon 
\beq
U=\pa{\ba{cc} 1 & i \\ i & 1  \ea} \qquad \e{and} \qquad 
U^{-1} = \f{1}{2} \pa{\ba{cc} 1 & -i \\ -i & 1  \ea} \; , 
\label{definition matrices U}
\enq
we get that the matrix
\beq\label{Msol}
M(\la) = D_{\infty}^{-\sg_3} \cdot U^{-1} \cdot \Big( \f{\la-a}{\la-b}\Big)^{ \f{\sg_3}{4} } \cdot  U \cdot D^{\sg_3}(\la)\,,
\enq
solves the Riemann--Hilbert problem
\begin{itemize}
\item $M \in \mc{O}\big(\Cx\setminus \intff{a}{b} \big)$ and has continuous boundary values
on $\intoo{a}{b}$;
\item $M(\la)=  \abs{ (\la-a)(\la-b) }^{-\f{1}{4}} \cdot  \e{O}\pa{ \ba{cc} 1 & 1 \\ 1 & 1 \ea}$ 
\;\; as $\la \tend a$ \;\; or \;\; $\la \tend b$; 
\item $M(\la) = I_{2} \; + \; \e{O}\big(\la^{-1}\big)$ when $\la \tend \infty$;
\item $M_-(\la) = M_+(\la) \cdot G_{M}(\la)$ for $\la \in \intoo{a}{b}$ where 
\beq
G_{M}(\la) = \pa{ \ba{cc} 0 &  - \ex{g(\la)} \\ 
						\ex{-g(\la)}  &  0 \ea }  \;. 
\label{MRH}
\enq
\end{itemize}

\subsection{Parametrix around $a$}\label{a}

Following \cite{DeiftItsZhouSineKernelOnUnionOfIntervals}, we construct the paramterix at $a$. 
We choose some $\de>0$ and define
\beq
E_a(\la) =  \f{\sqrt{\pi}\ex{i\f{\pi}{4}}}{2} M(\la) \ex{ \f{ g(\la) }{ 2 } \sg_3  } \f{1}{2} \pa{\ba{cc} 1 & -i \\ -i & 1 \ea}
\big[ - m^2 \zeta_a(\la) \big]^{- \f{ \sg_3 }{ 4 } }\,,
\enq
where 
\beq
\zeta_a(\la) = 
(\la-a) (b-\la)  \cdot \Bigg(  \Int{a}{b}  \f{p(\la)-p(s)}{\pi(\la-s)} 
\f{\dd s}{ (s - a)^{\f{1}{2}} \cdot (b - s)^{\f{1}{2}}  } \Bigg)^2\,.
\enq

In particular, one has that $\zeta_a^{\prime}(a)>0$, meaning that, at least for $\de$ small enough 
\beq
\zeta_a\big( \mc{D}_{a,\de} \cap \mathbb{H}_{\pm}\big) \subset \mathbb{H_{\pm}}\,, \qquad  \e{and} \qquad  
\inf_{\la \in \Dp{} \mc{D}_{a,\de} } | \zeta^{\prime}_a(\la) | \, > \, 0 \;. 
\enq
There $\mc{D}_{a,\de} = \big\{ z \in \Cx \; : \; |z-a|< \de  \big\}$ is the disc of radius $\de$
centred at $a$ and $\Dp{}\mc{D}_{a,\de}$ refers to its canonically oriented boundary. 
It then follows form the jump conditions satisfied by $M$ that 
$E_a$ is continuous across $\intoo{a}{a+\de}$.
Furthermore, the local behaviour of $M$ at $a$ ensures that the singularity of $E_a$ at $a$
is of removable type. As a consequence, we get that $E_a$ is
analytic on $\mc{D}_{a,\de}$.

Next, we introduce a matrix valued function built out of Hankel functions $H_a^{(k)}$ whose definition is recalled in Appendix \ref{Appendix}
\beq
\mc{P}_a^{(0)}(\la) = \pa{  \ba{cc}  
	\sqrt{ - m^2\zeta_a(\la) } \big[ H_0^{(2)}\big]^{\prime} \Big( \tf{ \sqrt{-m^2\zeta_a(\la)}}{2} \Big) &  
 \sqrt{ - m^2\zeta_a(\la) } \big[ H_0^{(1)}\big]^{\prime} \Big( \tf{  \sqrt{-m^2\zeta_a(\la) } }{ 2 } \Big) \\ 
\\
H_0^{(2)} \Big( \tf{ \sqrt{-m^2\zeta_a(\la)} }{ 2 } \Big)  &  H_0^{(1)}\Big( \tf{ \sqrt{-m^2\zeta_a(\la)} }{ 2 } \Big)  \ea }
\ex{  \big[ i \sqrt{-m^2\zeta_a(\la)} - g(\la) \big]  \f{\sg_3}{2} } \;. 
\enq
It follows from the asymptotic expansions of the Hankel functions \eqref{Appendix Asympt Exp Hankel 1}-\eqref{Appendix Asympt Exp Hankel prime 2}
that, uniformly in  $\la \in \Dp{}\mc{D}_{a,\de}$, 
\beq
\mc{P}_a^{(0)}(\la)  \simeq [- m^2\zeta_a(\la)]^{ \f{\sg_3}{4} } \f{2 \ex{-i\f{\pi}{4} }}{\sqrt{\pi}}
\pa{\ba{cc} 1 & i \\ i & 1 \ea } \sul{n \geq 0}{} (0,n) \Big(  \f{i}{\sqrt{-m^2\zeta_a(\la)}} \Big)^n
\pa{\ba{cc} (-1)^n a_n  &  i b_n \\
			-i (-1)^n b_n  & a_n  \ea}  \ex{  - g(\la)   \f{\sg_3}{2} }\,, 
\enq
where $\simeq$ indicates an equality in the sense of asymptotic expansions, while 
\beq
a_n = \f{1}{1-2n} \qquad ,  \qquad 
b_n = \f{2n}{1-2n}
\enq
and $(\nu,n)$ is as defined in \eqref{definition symbole nu n}. We are now in position to write down the parametrix at $a$:
\beq
\mc{P}_a(\la)= E_a(\la) \mc{P}_a^{(0)}(\la) \;. 
\enq
It satifies, uniformly in $\la \in \Dp{}\mc{D}_{a,\de}$, 
\beq
\mc{P}_a(\la) = M(\la) \cdot \Big[ I_2 + \e{O}(m^{-1}) \Big] \;,
\enq
uniformly on $\Dp{}\mc{D}_{a,\de}$. Elementary properties satisfied by the Hankel functions 
\eqref{Appendix ecriture saut Hankel 1}-\eqref{Appendix ecriture saut Hankel 1 prime} allow one to readily check 
that the parametrix $\mc{P}_a$ solves the Riemann--Hilbert problem
\begin{itemize}
\item $ \mc{P}_a \in \mc{O}\big(\mc{D}_{a,\de} \setminus \intfo{a}{a+\de} \big)$ and has continuous boundary values
on $\intoo{a}{a+\de}$;
\item $\mc{P}_a(\la)=  \ln|\la-a| \cdot  \e{O}\pa{ \ba{cc} 1 & 1 \\ 1 & 1 \ea}$ 
\; \; as  \; \; $\la \tend a$; 
\item $\mc{P}_a(\la) = M(\la) \cdot \Big[I_2 + \e{O}(m^{-1}) \Big]$ uniformly in  $\la \in \Dp{} \mc{D}_{a,\de}$;
\item $[\mc{P}_a]_-(\la) = [\mc{P}_a]_+(\la) \cdot  G_{\Xi}(z)$ for $\la \in \intoo{a}{a+\de}$. 
\end{itemize}
Also, it is easy to check that 
\beq
\mc{P}_a^{-1}(\la) \; \simeq \; \sul{n \geq 0}{} (0,n) \Big(  \f{i}{\sqrt{-m^2\zeta_a(\la)}} \Big)^n
\pa{\ba{cc}  a_n  &  - i b_n \ex{ g(\la) }  \\
			i (-1)^n b_n \ex{ -g(\la) }   & (-1)^n a_n  \ea}   M^{-1}(\la) \;. 
\enq

\subsection{Parametrix around $b$}\label{b}

The parametrix at $b$ is defined in a similar way. We introduce 
\beq
\zeta_b(\la) = 
(\la-a) (\la-b)  \cdot \Bigg(  \Int{a}{b}  \f{p(\la)-p(s)}{\pi(\la-s)} 
\f{\dd s}{ (s - a)^{\f{1}{2}} \cdot (b - s)^{\f{1}{2}}  } \Bigg)^2\,,
\enq
and then define 
\beq
E_b(\la) =  \f{\sqrt{\pi}\ex{i\f{\pi}{4}}}{2} M(\la) \ex{ \f{ g(\la) }{ 2 } \sg_3  } \f{1}{2} \pa{\ba{cc} 1 & -i \\ -i & 1 \ea}
\big[ m^2 \zeta_b(\la) \big]^{ \f{ \sg_3 }{ 4 } } \;. 
\enq

Again, one has that $\zeta_b^{\prime}(b)>0$, meaning that, at least for $\de$ small enough 
$\zeta\big( \mc{D}_{b,\de} \cap \mathbb{H}_{\pm}\big) \subset \mathbb{H_{\pm}}$. The same reasoning as before 
leads to the conclusion that $E_b\in \mc{O}( \mc{D}_{b,\de} )$. Further, the matrix 
\beq
\mc{P}_b^{(0)}(\la) \; = \;   \pa{  \ba{cc}  
H_0^{(1)}\Big( \tf{ \sqrt{m^2\zeta_b(\la)} }{ 2 } \Big)  & H_0^{(2)}\Big( \tf{ \sqrt{m^2\zeta_b(\la)} }{ 2 } \Big)   \\
\sqrt{ m^2\zeta_b(\la) } \big[ H_0^{(2)}\big]^{\prime}\Big( \tf{ \sqrt{m^2\zeta_b(\la)} }{ 2 } \Big) &  
 \sqrt{  m^2\zeta_b (\la) } \big[ H_0^{(1)}\big]^{\prime}\Big( \tf{ \sqrt{m^2\zeta_b(\la)} }{ 2 }  \Big)  
  \ea }
\ex{  - \big[  i \sqrt{m^2\zeta_b(\la)} \, + \,  g(\la) \big]  \f{\sg_3}{2} } \;, 
\enq
admits, uniformly in $\la \in \Dp{}\mc{D}_{b,\de}$, the asymptotic expansion
\beq
\mc{P}_b^{(0)}(\la) \; \simeq \;  \Big[ m^2\zeta_b(\la) \Big]^{ -\f{\sg_3}{4} } \f{2 \ex{-i\f{\pi}{4} }}{\sqrt{\pi}}
\pa{\ba{cc} 1 & i \\ i & 1 \ea } \sul{n \geq 0}{} (0,n) \Big(  \f{i}{\sqrt{ m^2\zeta_b(\la)}} \Big)^n
\pa{\ba{cc} a_n  &  -i (-1)^n  b_n \\
			i  b_n  &  (-1)^n  a_n  \ea}  \ex{  - g(\la)   \f{\sg_3}{2} } \;. 
\enq
 The full parametrix in the neighbourhood of $b$ then reads 
\beq
\mc{P}_b(\la)= E_b(\la) \mc{P}_b^{(0)}(\la) \;. 
\enq
The latter solves the Riemann--Hilbert problem
\begin{itemize}
\item $ \mc{P}_b \in \mc{O}\big(\mc{D}_{b,\de} \setminus \intof{b-\de}{b} \big)$ and has continuous boundary values
on $\intoo{b-\de}{b}$;
\item $\mc{P}_b(\la)=  \ln|\la-b| \cdot  \e{O}\pa{ \ba{cc} 1 & 1 \\ 1 & 1 \ea}$ 
\; \; as \;\;  $\la \tend b$; 
\item $\mc{P}_b(\la) = M(\la) \cdot \Big[ I_2 + \e{O}(m^{-1}) \Big]$ uniformly in  $z \in \Dp{} \mc{D}_{b,\de}$;
\item $[\mc{P}_b]_-(\la) = [\mc{P}_b]_+(\la) \cdot  G_{\Xi}(\la)$ for $\la \in \intoo{b-\de}{b}$. 
\end{itemize}

Finally, one can check that 
\beq
\mc{P}_b^{-1}(\la) \; \simeq  \; \sul{n \geq 0}{} (0,n) \Big(  \f{i}{\sqrt{m^2\zeta_b(\la)}} \Big)^n
\pa{\ba{cc}  (-1)^n a_n  &   i (-1)^n b_n \ex{ g(\la) }  \\
			- i  b_n \ex{ -g(\la) }   &  a_n  \ea}   M^{-1}(\la) \;. 
\enq

\subsection{The last transformation: the matrix $\Ups$}
\label{SousSectionRHP for Ups}

We define a piecewise analytic matrix $\Ups$ by
\beq\label{Upsdef}
\Ups(\la) = \left\{ \ba{cc}    
\Xi(\la) \cdot M^{-1}(\la)\,, &  \la \in  \Cx \setminus  \Big\{  \Dp{} \mc{D}_{a,\de} \cup \Dp{} \mc{D}_{b,\de} \cup
\intff{a+\de}{b-\de} \Big\}\,, \\
\Xi(\la) \cdot \mc{P}_a^{-1}(\la)\,, &  \la \in   \mc{D}_{a,\de}\,, \\
\Xi(\la) \cdot \mc{P}_b^{-1}(\la)\,, &  \la \in  \mc{D}_{b,\de}\,.     
\ea \right. 
\enq

It is then readily checked that the matrix $\Ups$ has its jumps solely on the two disks $-\Dp{}\mc{D}_{a,\de}$
and $-\Dp{}\mc{D}_{b,\de}$ endowed with an opposite (in respect to the canonical one) orientation. Namely,
the matrix $\Ups$ is the unique solution to the Riemann--Hilbert problem
\begin{itemize}
\item $ \Ups \in \mc{O}\Big(\Cx \setminus  \Dp{}\big\{  \mc{D}_{a,\de} \cup \mc{D}_{b,\de} \big\} \Big)$ and has continuous boundary values
on $-\Dp{}\mc{D}_{a,\de} \cup -\Dp{}\mc{D}_{b,\de}$;
\item $\Ups(\la) =  I_2 \; + \;  \e{O}(\la^{-1})$ when $\la \tend \infty$;
\item $\Ups_-(\la) = \Ups_+(\la) \cdot G_{\Ups}(\la)$ for $z \in -\Dp{}\mc{D}_{a,\de} \cup -\Dp{}\mc{D}_{b,\de}$, where 
\beq
G_{\Ups}(\la) = M(\la) \cdot \mc{P}_a^{-1}(\la) \quad \e{for} \quad \la \in -\Dp{}\mc{D}_{a,\de} \qquad \e{and} \qquad
 M(\la) \cdot \mc{P}_b^{-1}(\la) \quad \e{for} \quad \la \in -\Dp{}\mc{D}_{b,\de}   \;. 
\enq
\end{itemize}

\subsubsection{Asymptotic expansion of the jump matrix $G_{\Ups}$ }

In the neighbourhood of $a$, it is readily seen that the matrix $G_{\Ups}(\la)$ admits the asymptotic
expansion
\beq
G_{\Ups}(\la) \simeq  \sul{n \geq 0}{}  (0,n)  \Big( \f{i}{ \sqrt{-m^2\zeta_a(\la) } } \Big)^n
M(\la)  \pa{\ba{cc}  a_n & -i b_n \ex{g(\la)} \\
						i(-1)^n b_n   \ex{-g(\la)}  & (-1)^n a_n \ea} M^{-1}(\la) \;. 
\enq
We then introduce 
\beq
\be(\la) = \ex{g(\la)} D^{2}(\la)  \qquad , \qquad r(\la) = \f{ \be(\la) \, + \, \be^{-1}(\la) }{ 2} \qquad \e{and} \qquad 
t(\la) = \f{ \be(\la) \, - \,  \be^{-1}(\la) }{ 2 q(\la)} \;.
\enq
The functions $r$ and $t$ are holomorphic on $\mc{D}_{a,\de}\cup\mc{D}_{b,\de}$

Then, agreeing upon 
\beq
u(\la) = \Int{a}{b} \f{p(\la)-p(s) }{\pi (\la-s) } \f{\dd s }{ (s-a)^{\f{1}{2}}(b-s)^{\f{1}{2}} }\,,
\enq
one has the identities
\beq
\f{ \be(\la)-\be^{-1}(\la) }{2\sqrt{-\zeta_a(\la)}} = \f{  t(\la) }{ u(\la) } \qquad 
  \Big( \f{\la-a}{ \la-b} \Big)^{\f{1}{2}} \cdot  \f{ \be(\la)-\be^{-1}(\la) }{2} =  t(\la) (\la-a)
\qquad \e{and} \qquad 
\Big( \f{ \la-b}{\la-a}  \Big)^{\f{1}{2}} \cdot  \f{ \be(\la)-\be^{-1}(\la) }{2} =  t(\la) (\la-b)\,,
\enq
as well as
\beq
  \Big( \f{\la-a}{ \la-b} \Big)^{\f{1}{2}} \cdot  \f{ 1 }{ \sqrt{-\zeta_a(\la)} } = \f{1}{(\la-b) u(\la)} 
\qquad \e{and} \qquad
  \Big( \f{ \la - b }{ \la - a } \Big)^{\f{1}{2}} \cdot  \f{ 1 }{ \sqrt{-\zeta_a(\la)} } = \f{1}{(\la-a) u(\la)} \;. 
\enq
What leads, after a quick computation, to 
\bem
U D_{\infty}^{\sg_3} G_{\Ups}(\la) D_{\infty}^{-\sg_3}   U^{-1}  = I_2 \; + \; \sul{p=1}{+\infty}
\f{ (0,2p) }{ m^{2p} \zeta_a^{p}(\la) }  
		\pa{  \ba{cc} a_{2p}-b_{2p}r(\la) &  - i (\la-a)b_{2p} t(\la)  \\ 
				-i (\la - b )b_{2p} t(\la) & a_{2p} + b_{2p} r(\la) \ea}   \\
\; + \;  \sul{p=0}{+\infty}
\f{ i (0,2p+1) }{ m^{2p+1} \zeta_a^{p}(\la) }  
		\pa{  \ba{cc} -b_{2p+1}\tf{ t(\la) }{ u(\la) } &  - i \tf{ \big[ a_{2p+1} + b_{2p+1} r(\la) \big] }{ \big[ (\la-b) u(\la) \big]}   \\ 
				i \tf{ \big[ a_{2p+1} - b_{2p+1} r(\la) \big] }{ \big[ (\la-a) u(\la) \big] }  &  b_{2p+1} \tf{ t(\la) }{ u(\la) } \ea}\,.  
\end{multline}

In particular, we get that, to the first order 
\beq
G_{\Ups}(\la) \simeq I_2 \; + \; \f{ G_1^{(a)}(\la) }{ m }  \; + \; \e{O}( m^{-2} ) \;, 
\label{ecriture DA GUps en a}
\enq
with 
\beq
G_1^{(a)}(\la) = - \f{ (0,1) (2r(\la)-1)  }{(\la-a) u(\la)} D_{\infty}^{-\sg_3} U^{-1} \sg^-   U  D_{\infty}^{\sg_3}
\; + \; G_{1;\e{reg}}^{(a)}(\la) \;. 
\enq
There $G_{1;\e{reg}}^{(a)}(\la)$ is a holomorphic matrix in some sufficiently small neighbourhood of $a$.

Likewise, in the neighbourhood of $b$ one has the asymptotic expansion
\bem
U D_{\infty}^{\sg_3} G_{\Ups}(\la) D_{\infty}^{-\sg_3}   U^{-1}  \simeq   I_2 \; + \; \sul{p=1}{+\infty}
\f{1}{ m^{2p} (\la-b)^p }
\f{ (-1)^p (0,2p) }{  (\la-a)^p u^{2p}(\la) }  
		\pa{  \ba{cc} a_{2p}+b_{2p}r(\la) &   i (\la-a)b_{2p} t(\la)  \\ 
				i (\la - b )b_{2p} t(\la) & a_{2p} - b_{2p} r(\la) \ea}   \\
\; + \;  \sul{p=0}{+\infty} \f{1}{ m^{2p+1} (\la-b)^p }
\f{ i (-1)^p (0,2p+1) }{ (\la-a)^{p} u^{2p+1}(\la) }  
		\pa{  \ba{cc} -b_{2p+1} t(\la)  &   i  \big[ a_{2p+1} - b_{2p+1} r(\la) \big] \cdot  (\la-b)^{-1}    \\ 
				-i  \big[ a_{2p+1} + b_{2p+1} r(\la) \big] \cdot  (\la-a)^{-1}   &  b_{2p+1} t(\la)  \ea}   \;. 
\end{multline}

Similarly to \eqref{ecriture DA GUps en a}, one has 
\beq
G_{\Ups}(\la) \; = \;  I_2 \; + \; \f{ G_1^{(b)}(\la) }{ m }  \; + \; \e{O}( m^{-2} ) \;, 
\enq
with 
\beq
G_1^{(b)}(\la) = - (0,1) \f{ (2 r(\la)-1)  }{(\la-b) u(\la)}  D_{\infty}^{-\sg_3} U^{-1} \sg^+   U  D_{\infty}^{\sg_3}
\; + \; G_{1;\e{reg}}^{(b)}(\la) \;. 
\enq
Here, again, $G_{1;\e{reg}}^{(b)}(\la)$ is a holomorphic matrix on some sufficiently small neighbourhood of $b$.

\subsubsection{Asymptotic resolution of the Riemann--Hilbert problem for $\Ups$}

The Riemann--Hilbert problem for $\Ups$ is equivalent to the singular integral equation \cite{BealsCoifmanScatteringInFirstOrderSystemsEquivalenceRHPSingIntEqnMention}
\beq
\Ups_{+}(\la) = I_2 \; + \; \Int{\Sg_{\Ups} }{} \Ups_{+}(s)  \f{ G_{\Ups}(s)-I_2 }{ (\la_+ - s) }  \cdot \f{ \dd s }{ 2i\pi  } \;, \qquad 
\Sg_{\Ups} = \Big\{ - \Dp{}\mc{D}_{a,\de} \Big\}\cup \Big\{ - \Dp{}\mc{D}_{b,\de} \Big\}\,,
\enq
which, due to the bound $G_{\Ups}-I_2 = \e{O}\big(m^{-1}\big)$ which holds in $\big(L^1\cap L^{\infty}\big)\big( \Sg_{\Ups} \big)$, can be 
solved in terms of a Neumann series \cite{CalderonContinuityCauchyTransformLipschitzCurves}. Recalling that the solution to the Riemann--Hilbert problem stated in 
Subsection \ref{SousSectionRHP for Ups} admits the representation 
\beq
\Ups(\la) = I_2 \; + \; \Int{\Sg_{\Ups} }{} \Ups_{+}(s)  \f{ G_{\Ups}(s)-I_2 }{ (\la - s) }  \cdot \f{ \dd s }{ 2i\pi  }\,,
\enq
it follows that, uniformly away from $\mc{D}_{a,\de}\cup\mc{D}_{b,\de}$,
\beq
\Ups(\la) = I_2 + \f{ (0,1 )}{ m } D_{\infty}^{ - \sg_3}  U^{-1} 
\bigg\{ \f{\sg^-}{(\la-a)u(a)} \; + \;  \f{\sg^+}{(\la-b)u(b)}  \bigg\} U D_{\infty}^{\sg_3} 
\;\; + \;\; \e{O}\big(m^{-2}\big) \;, 
\enq
with a remainder that holds in the $L^{\infty}\big( \Cx \setminus \ov{\mc{D}}_{a,2\de} \cup \ov{\mc{D}}_{b,2\de}  \big)$ sense since 
$G_{\Ups}$is holomorphic in the neighbourhood of $\Sg_{\Ups}$. 

In view of the definition (\ref{Upsdef}), it holds 
\begin{equation}\label{Xias}
\Xi(\lambda) =
 \bigg(  
 I_2 + \f{ (0,1 )}{ m } D_{\infty}^{ - \sg_3}  U^{-1} 
\bigg\{ \f{\sg^-}{(\la-a)u(a)} \; + \;  \f{\sg^+}{(\la-b)u(b)}  \bigg\} U D_{\infty}^{\sg_3} 
\;\; + \;\; \e{O}\big(m^{-2}\big) 
\bigg) \cdot M(\la),
\end{equation}
where the $\e{O}\big(m^{-2}\big) $ remainder holds in the $L^{\infty}\big( \Cx \setminus \ov{\mc{D}}_{a,2\de} \cup \ov{\mc{D}}_{b,2\de}  \big)$ sense. 

Thus, one is led to 
\beq \label{chifinal}
 \chi(\la) \; = \; \ex{- m h_{\infty} \sg_3 } \bigg\{  
 I_2 + \f{ (0,1 )}{ m } D_{\infty}^{ - \sg_3}  U^{-1} 
\bigg\{ \f{\sg^-}{(\la-a)u(a)} \; + \;  \f{\sg^+}{(\la-b)u(b)}  \bigg\} U D_{\infty}^{\sg_3} 
\;\; + \;\; \e{O}\big(m^{-2}\big) 
\bigg\} \cdot M(\la) \cdot \ex{ m h (\la)  \sg_3 }  \;, 
\enq
for $\la \in \Cx \setminus \ov{\mc{D}}_{a,2\de} \cup \ov{\mc{D}}_{b,2\de}$.

\section{The determinant}
\label{Section asymptotic evaluation of the determinant}

In the present section, we build on the asymptotic expansion obtained in the previous section so as to obtain the leading asymptotics
of $\det[I+V]$. Our proof builds on certain differential identities that have been first established in 
\cite{KozKitMailSlaTerRHPapproachtoSuperSineKernel}.

\subsection{Differential identities for the determinant}

Let the function $p$ arising in the definition of the kernel \eqref{definition noyau GSK}
depend smoothly on an auxiliary variable $t$: $p(\la) \hookrightarrow p(\la,t)$, this in such a way that 
$\la \mapsto \Dp{t}p(\la,t) \in \mc{O}(U)$ uniformly in $t \in \intff{0}{1}$. 
Let $V_t$ denote the associated kernel and $\chi(\la,t)$ the solution to the corresponding Riemann--Hilbert problem. 

Then, by repeating the handlings developed in \cite{KozKitMailSlaTerRHPapproachtoSuperSineKernel}, one obtains the representation
\beq
\Dp{t} \ln \ddet{}{I+V_t} \; = \; m \Oint{ \Ga }{} \e{tr} \Big[ \Dp{\la}\chi(\la,t) \cdot \sg_3 
\cdot \chi^{-1}(\la,t) \Big] \cdot \f{ \Dp{t}p(\la,t) }{4\pi }  \cdot \dd \la \;. 
\label{formule derivee t determinant}
\enq
Above $\Ga$ refers to a small counterclockwise loop around $\intff{a}{b}$ that lies in $U$.

%
%
%
%
%
%
%
%
%

\subsection{Asymptotic behaviour of $\ddet{}{I+V}$}

\begin{prop}\label{prop1}
The large-$m$ asymptotics of the Fredholm determinant associated with the integral operator $I+V$
takes the form 
\bem\label{detgen}
\ln \ddet{}{I \, + \, V} \; = \;  m^2  \Int{a}{b}  \f{ 2\la \mu \, + \,  2 ab \,- \, (a+b)(\la + \mu)  }
{ \sqrt{ (\la-a)(b-\la) (\mu-a)(b-\mu) } }   
\cdot   \bigg(  \f{ p(\la) -p(\mu) - i \tf{ [g(\la)-g(\mu)] }{ m }  }{  \la - \mu  } \bigg)^2 
 \cdot   \f{ \dd \la \dd \mu }{ (4\pi)^2 }  \\
 -  \f{1}{4} \ln m  + \f{1}{8} \ln \bigg(  \f{16 u(a)u(b)}{(b-a)^2}  \bigg)   \; + \; \f{ \ln 2}{12} 
 \; + \; 3 \zeta^{\prime}(-1)\;+ \;  \e{O}\big( m^{-1} \big) \;. 
\end{multline}
\end{prop}

\Proof 

In order to get the leading asymptotic behaviour of the determinant one can, for the purpose of the intermediate calculations\footnote{It is of 
course not a problem to keep $g\not=0$ throughout all of the below calculation.} set $g=0$. In such a setting, one has $D=0$. The whole dependence on $g$ in the leading asymptotics 
can then be recovered from the replacement $ p \hookrightarrow  p - i \tf{g}{m}$. 
We shall build on the representation \eqref{formule derivee t determinant} so as to interpolate, in the large-$m$
regime, between the large-$m$ asymptotics of a known case -the pure sine kernel 
which correspond to $p(\la)=\la$- and our case of general holomorphic $p$. Thus, we introduce 
\beq
\la \mapsto p(\la,t) \; = \; t p(\la) \; + \; (1-t) \la \;, 
\enq
This function is holomorphic in the open neighbourhood $U$ of $\intff{a}{b}$ and it further satisfies 
\beq
\Dp{\la} p(\la,t) \, > \, \min( 1, \inf_{\intff{a}{b}}\Dp{\la}p(\la) ) \;, 
\enq
all this uniformly in $t\in \intff{0}{1}$. As a consequence, the previous analysis can be carried out for any
value of $t\in \intff{0}{1}$. Furthermore, all the remainders will be uniform in $0 \leq t \leq 1$. 
Note that dealing with a $t$-dependent function $p$ will generate an additional $t$-dependence in the auxiliary functions 
$h$ and $u$, \textit{viz}.: 
\beq
h(\la,t) \; = \; q(\la) \Int{a}{b} \f{ -ip(\la,t) }{q_{+}(s) (s-\la) } \cdot \f{ \dd s }{ 2i\pi} \qquad \e{and} \qquad
u(\la,t) \; = \;  \Int{a}{b} \f{ p(\la,t)- p(s,t) }{\pi (s-\la) } \cdot \f{ \dd s }{ \sqrt{(s-a)(b-s) } } \;. 
\enq
In order to calculate \textit{rhs} of \eqref{formule derivee t determinant} asymptotically in $m$, we first recast 
\beq
\e{tr}\big[ \Dp{\la}\chi(\la,t) \cdot \sg_3 \cdot \chi^{-1}(\la,t) \big]
\enq
in terms of the objects arising in the various steps of the transformations carried out on the initial RHP for $\chi$. 

Using that, uniformly away from the endpoints $a,b$, one has
\beq
 \chi(\la,t) \; = \; \ex{- m h_{\infty}(t) \sg_3 }\cdot \Ups(\la,t) \cdot M(\la) \cdot \ex{ m h (\la,t)  \sg_3 }  \;,   
\enq
we are led to 
\bem
\e{tr}\Big[ \Dp{\la}\chi(\la,t) \cdot \sg_3 \cdot \chi^{-1}(\la,t) \Big] \; = \; 2m \Dp{ \la }h(\la,t) 
\; + \; \e{tr}\Big[ \Dp{\la}M(\la) \cdot \sg_3 \cdot M^{-1}(\la) \Big] \\
\; + \; \e{tr}\Big[ \big( \Dp{\la} \Ups(\la,t) \big) \cdot M(\la) \cdot \sg_3 \cdot M^{-1}(\la) \cdot \Ups(\la,t) \Big]  \;. 
\end{multline}
Furthermore, as a consequence of 
\beq
 U \sg_3 U^{-1} \; = \; \left( \ba{cc}  0 & -i \\
 										i & 0  \ea  \right) \;. 
\enq
one gets that 
\beq
\e{tr}\Big[ \Dp{\la}M(\la) \cdot \sg_3 \cdot M^{-1}(\la) \Big]  \; = \; \f{1}{4} \bigg\{ \f{1}{\la - a}  - \f{1}{\la - b} \bigg\} 
\cdot \e{tr}\Big[ U^{-1} \sg_3 U \sg_3 \Big]  \; = \; 0 \;. 
\enq
Finally, the leading in $m$ asymptotics of $\Ups$
\beq
\Ups(\la,t) \; = \; I_2 + \f{ (0,1 )}{ m }   U^{-1} 
\bigg\{ \f{\sg^-}{(\la-a)u(a,t)} \; + \;  \f{\sg^+}{(\la-b)u(b,t)}  \bigg\} U  
\;\; + \;\; \e{O}\big(m^{-2}\big) \,,
\enq
yield
\bem
 \e{tr}\Big[ \big( \Dp{\la} \Ups(\la,t) \big) \cdot M(\la) \cdot \sg_3 \cdot M^{-1}(\la) \cdot \Ups(\la,t) \Big]   
%
%
\\
\; = \;- \f{i (0,1) }{ m } \cdot \Bigg[ \f{1}{(\la-b)^\f{3}{2} (\la-a)^{\f{1}{2}} u(b,t)}  
\; - \; \f{ 1 }{(\la-a)^\f{3}{2} (\la-b)^{\f{1}{2}} u(a,t)}  \Bigg] \;\; + \;\; \e{O}\big(m^{-2}\big)  \;. 
\end{multline}

As a consequence, we are led to the representation 
\beq
\Dp{t}\ln \det[I\, +\, V_t] \; = \; 
\f{2m^2}{4\pi} \Oint{ \Ga }{} \Dp{t}p(\la , t ) \cdot \Dp{\la}h(\la,t) \cdot \dd \la 
\; + \; C_t \; + \; \e{O}\big(\f{1}{m} \big)  \;, 
\label{ecriture asympt dom derivee en t du det}
\enq
where the remainder is uniform in $t\in \intff{0}{1}$ and the constant $C_t$ reads 
\beq
C_t \; = \;    \f{ i(0,1) }{ 4\pi } \Oint{ \Ga   }{} \bigg\{   
\f{\Dp{t}p(\la, t) }{ u(a,t)(\la-a)^\f{3}{2} (\la-b)^{\f{1}{2}} }   \; - \;  
 \f{\Dp{t}p(\la, t) }{ u(b,t)(\la-b)^\f{3}{2} (\la-a)^{\f{1}{2}} }\bigg\}\cdot \dd \la \;. 
\label{definition constante Ct}
\enq
The constant $C_t$ can be estimated as follows. 
Since the residue at $\infty$ does not contribute, one can regularise the behaviour of the integrand 
in the vicinities of $a$ and $b$ leading to 
\bem
C_t \; = \; - \f{ i(0,1) }{ 2\pi u(a,  t ) } \Int{ a  }{ b } 
\f{  \Dp{t}p(\la , t)- \Dp{t}p(a, t) }{ (\la-a)^\f{3}{2} (\la-b)^{\f{1}{2}}_+ } \cdot \dd \la  
\; + \;    \f{ i(0,1) }{ 2\pi u(b, t) } \Int{ a  }{ b } 
\f{  \Dp{t}p(\la, t)- \Dp{t}p(b, t) }{ (\la-a)^\f{1}{2} (\la-b)^{\f{3}{2}}_+ } \cdot \dd \la  \\
\; = \;   \f{1}{8 u(a ,  t) }  \Dp{t}u(a ,  t)  \;  + \; \f{1}{8 u(b ,  t) }  \Dp{t}u(b,  t)  \; = \; 
 \f{1}{8} \Dp{t} \ln \big[u(a , t) u(b , t) \big]   \;. 
\end{multline}
Thus, it solely remains to recast the first term in \eqref{ecriture asympt dom derivee en t du det}. 
Starting with the integral representation \eqref{ecriture rep int h pour la exterieur ctr}
for the function $h$ and upon taking the derivatives explicitly, we get 
\beq
 \Oint{ \Ga }{} \Dp{t}p(\la , t ) \cdot \Dp{\la}h(\la,t) \cdot \f{ \dd \la }{2\pi}  \; = \; 
%
%
%
%
%
%
%
%
%
%
\f{1}{16\pi^2}  \Oint{ \Ga }{}   \dd \la \Oint{ \Ga^{\prime} }{}  \dd s 
\f{ \Dp{t}p(\la , t ) \cdot p(s, t) }{ (s-\la)^2 \cdot q(\la) \cdot q(s)    } \cdot 
\Big[ 2 \la s  \; + \;  2 a b  \; - \; (a+b)(\la+s) \Big]\,,
\enq
where we agree that $\Ga$ and $\Ga^{\prime}$ are two loops around $\intff{a}{b}$ such that 
$\Ga^{\prime} \subset \Ga $.  
We could have of course started with the representation \eqref{ecriture rep int h pour la int ctr} 
for $h$. Then, the part involving $-i p(\la,t)/2$ produces a vanishing contribution by squeezing the contour $\Ga$
to $0$. Further, one may get rid of the contribution to $h(\la,t)$ involving  $p(\la, t)$ under the integral sign 
by deforming the $s$-contour in \eqref{ecriture rep int h pour la int ctr}  to infinity. Then, one is led 
to exactly the same integral representation as above, with the sole difference that the
contours $\Ga$ and $\Ga^{\prime}$ are interchanged, \textit{i.e.}
\beq
 \Oint{ \Ga }{} \Dp{t}p(\la , t ) \cdot \Dp{\la}h(\la,t) \cdot \f{ \dd \la }{2\pi}  \; = \; 
\f{1}{16\pi^2}  \Oint{ \Ga^{\prime} }{}   \dd \la \Oint{ \Ga }{}  \dd s 
\f{ \Dp{t}p(\la , t ) \cdot p(s , t) }{ (s-\la)^2 \cdot q(\la) \cdot q(s)    } \cdot 
\Big[ 2 \la s  \; + \;  2 a b  \; - \; (a+b)(\la+s) \Big] \;. 
\enq
Hence, we get that 
\bem
 \Oint{ \Ga }{} \Dp{t}p(\la , t ) \cdot \Dp{\la}h(\la,t) \cdot \f{ \dd \la }{2\pi}  \; = \; 
\f{1}{32\pi^2}  \f{\Dp{}}{\Dp{}t} \Oint{ \Ga }{}   \dd \la \Oint{ \Ga^{\prime} }{}  \dd s 
\f{ p(\la , t ) \cdot p(s , t) }{ (s-\la)^2 \cdot q(\la) \cdot q(s)    } \cdot 
\Big[ 2 \la s  \; + \;  2 a b  \; - \; (a+b)(\la+s) \Big]  \\
\; = \; -\f{1}{64\pi^2}  \f{\Dp{}}{\Dp{}t} \Oint{ \Ga }{}   \dd \la \Oint{ \Ga^{\prime} }{}  \dd s 
\f{ 2 \la s  \; + \;  2 a b  \; - \; (a+b)(\la+s)  }{  q(\la) \cdot q(s)    } \cdot 
\bigg\{  \Big( \f{ p(s , t) - p(\la ,  t) }{s-\la} \Big)^2 \; - \; 
\f{ p^2(s , t) + p^2(\la , t) }{ (s-\la)^2 }  \bigg\} \\
\; = \; -\f{1}{16\pi^2}  \f{\Dp{}}{\Dp{}t} \Int{ a }{b}  
\f{ 2 \la s  \; + \;  2 a b  \; - \; (a+b)(\la+s)  }{  q_+(\la) \cdot  q_+(s)    } \cdot 
 \Big( \f{ p(s , t) - p(\la , t) }{s-\la} \Big)^2 \cdot  \dd \la \dd s  
 \; + \; G \;,  
\end{multline}
The first part is already of the desired form, whereas $G$ is given by
\beq
G \; = \;  \f{1}{64\pi^2}  \f{\Dp{}}{\Dp{}t} \Oint{ \Ga }{}   \dd \la \Oint{ \Ga^{\prime} }{}  \dd s 
\f{ 2 \la s  \; + \;  2 a b  \; - \; (a+b)(\la+s)  }{   q(\la) \cdot q(s)   \cdot(s-\la)^2   } 
\big[ p^2(s , t) + p^2(\la , t) \big ] \;. 
\enq

Due to the inclusion of contours $\Ga^{\prime} \supset \Ga$, the integral involving $p^2(\la, t) $
yields $0$ by taking the residue in the $s$-integral at infinity. Now, in order to estimate the integral involving 
$p^2(s , t) $, we deform the contour $\Ga$ up to $\infty$. Again the residue at $\infty$ does not contribute, but we have to take into 
account the residue at $\la = s$. All of these manipulations recast $G$ as
\beq
G \; = \;  \f{- i}{32\pi}  \f{\Dp{}}{\Dp{}t} \Oint{ \Ga }{}   \dd s 
\bigg\{ 2s -a -b - \f{1}{2} \Big( \f{1}{s-a} + \f{1}{s-b}  \Big) 2 (s^2-s(a+b) + ab) \bigg\}
\f{ p^2(s , t) }{ q^2(s) }  \; = \; 0 \;. 
\enq
All in all, by putting all the pieces of the analysis together, we are led to the representation 
\bem
\Dp{t} \ln \det[I+V_t] \; = \; 
 \f{ m^2 }{16\pi^2}  \f{\Dp{}}{\Dp{}t} \Int{ a }{b}   
\f{ 2 \la s  \; + \;  2 a b  \; - \; (a+b)(\la+s)  }{ \sqrt{ (\la-a)(b-\la)(s-a)(b-s) }    }\cdot \bigg( \f{ p(s,t)-p(\la,t) }{ s-\la} \bigg)^2 \cdot \dd \la \dd s  \\
\; + \; \f{1}{8} \f{ \Dp{} }{ \Dp{} t} \ln \big[u(a , t) u(b , t) \big]   \; + \; \e{O}\big( m^{-1} \big)\,.
\end{multline}
Thus integrating from $t=0$ up to $t=1$,  inserting the large-$m$ asymptotic behaviour of the pure sine kernel \cite{dik,EhrhardtConstantinPureSinekernelFredholmDeyt,kras}
\beq
\ln \det[I+V_0] \; = \; - \f{ (b-a)^2 }{32}m^2 \; - \; \f{1}{4} \ln \Big[ \f{b-a}{4} \cdot m \Big] \; + \; \f{ \ln 2 }{ 12 } \; + \; 
3\zeta^{\prime}(-1) \; + \; \e{O}\big( m^{-1} \big) \; ,
\enq
and using that 
\beq
\Int{a}{b}  \f{ \dd s }{\pi \sqrt{ (s-a)(b-s)} } \; = \; 1 \qquad \e{along} \; \e{with} \qquad 
\Int{ a }{b}  \f{ 2 \la s  \; + \;  2 a b  \; - \; (a+b)(\la+s)  }{ \sqrt{ (\la-a)(b-\la)(s-a)(b-s) }    }
\cdot  \dd \la \dd s  \; = \; - \f{\pi^2}{2} (b-a)^2 \;, 
\enq
we are led to the claim. \qed





\section{Emptiness formation probability in the XXZ-spin 1/2 Heisenberg chain.}\label{efp}

In this section, as it has already been indicated in the introduction,  we will consider one special example of the generalised sine kernel which has a particular
importance in the theory of the XXZ spin-1/2 Heisenberg chain. 

Let $V $ be the trace class  integral operator acting on $L^2(\Gamma_{\alpha})$, where $\Gamma_{\alpha}$ 
is the arc,
$$
|\lambda| = 1, \quad -\alpha< \arg \lambda < \alpha \quad 0<\alpha < \pi,
$$
traversed counterclockwise. The operator's kernel is given by
 \begin{equation}\label{010}
	V(\lambda,\mu)=-\frac{1}{2 \pi i (\lambda - \mu)} \left( \di \lambda^{m/2} \mu^{-m/2} e^{t(\frac{\phi(\lambda)-\phi(\mu)}{2})} - \lambda^{-m/2} \mu^{m/2} e^{t(\frac{\phi(\mu)-\phi(\lambda)}{2})} \right)
\end{equation}
where, as before, $m$ is a positive integer, $t$ is a real parameter, and  the function $\phi(\lambda)$ is
assumed to be analytic in the neighbourhood of the arc $\Gamma_{\alpha}$. In the following,  we explain the connection
of this kernel  to  the XXZ spin-1/2 Heisinberg chain.

The $XXZ$ spin-1/2 Heisenberg chain of size $N$ is determined by the Hamiltonian,
\begin{equation}\label{xxz}
H_{XXZ}= \sum_{n= 1}^{N}
\left(\sigma^x_{n}\sigma^x_{n+1}+\sigma^y_{n}\sigma^y_{n+1} + \Delta \left(\sigma^z_{n}\sigma^z_{n+1} -1\right) -h\sigma^z_n\right),
\end{equation}
where the periodic boundary conditions are assumed. In (\ref{xxz}),
$\sigma^x$, $\sigma^y$, $\sigma^z$ are Pauli matrices,
$h$, $0\leq h < 4(1+\De)$,  is an external (moderate) magnetic field, and  $\Delta$ is the anisotropy parameter which takes the values  $-1 < \Delta < 1$. 
At the point $\Delta = 0$, the model becomes the free fermionic XX0 spin chain.

One of the principal objects of the analysis of the XXZ model is the  {\it emptiness formation probability} (EFP) which
is defined at zero temperature  as the correlation function,
\begin{equation}\label{Pn}
P^{(N)}(m) = \langle \boldsymbol{\psi}_g, \, \prod_{j=1}^{m}\frac{\sigma_j^z + 1}{2}  \boldsymbol{\psi}_g \rangle \, .
\end{equation}
The physical meaning of $P^{(N)}(m)$ is   the probability of finding a string of $m$ adjacent  parallel  spins up
(i.e., a piece of the {\it ferromagnetic } state) in the antiferromagnetic ground 
state $ \boldsymbol{\psi}_g$ for a given value of the magnetic field $h$. We shall denote,
\begin{equation}\label{Pninf}
P(m):= \lim P^{(N)}(m), \quad N \to \infty,
\end{equation}
the emptiness formation probability in the thermodynamic limit. The existence of this limit follows from the works 
\cite{KitanineMailletTerrasElementaryBlocksPeriodicXXZ,KozProofOfDensityOfBetheRoots}. The principal analytical question is the large $m$ behaviour
of $P(m)$. 

At the free fermonic case, when $\Delta =0$, the EFP is given by the explicit determinant formula involving the integral
operator (\ref{010}). Indeed,  one has that
\begin{equation}\label{freefer}
P(m)_{\mid \De=0} = \det\Bigl[I + V\Bigr]\Bigl|_{t=0} \, .
\end{equation}
The Fredholm determinant in the right hand side of this formula can also be expressed as a Toeplitz determinant whose symbol
is the characteristic function of the complimentary arc, $C\setminus \Gamma_{\alpha}$. The large $m$ asymptotics of this
determinant was obtained in the classical work by Widom  \cite{w}  and it reads (see also \cite{dik,kras} for the error estimate),
\begin{equation}\label{wdet}
P(m)  = m^2\ln\cos{\frac{\alpha}{2}} 
 - \frac{1}{4} \ln \left(m\sin{\frac{\alpha}{2}}\right)
+ c_0 + O\left(\frac{1}{m}\right),\quad m \to \infty \, ,
\end{equation}
where the constant $c_0$ is the famous Widom's constant 
$$
c_0 = \frac{1}{12}\ln 2 + 3\zeta'(-1).
$$

There exists a Fredholm determinant representation for the EFP in the general XXZ case
as well. A remarkable fact is that this representation also involves  the operator $V$ but this
time with $t \neq 0$. The exact formula relating $P(m)$ and $V$ for $\Delta \neq 0$
was extracted  by N. Slavnov by using the results obtained  in \cite{KozKitMailSlaTerXXZsgZsgZAsymptotics}.   
The  function $\phi(\lambda)$ in Slavnov's formula, however, is not a scalar function, but is in fact
a {\it dual quantum field}  acting in an auxiliary bosonic  Fock space with vacuum
$|0\rangle$. Indeed, Slavnov's representation takes the form
\begin{equation}\label{PnFrednonfree}
 P(m)=\langle 0 \big|{
 C(\varphi)}\cdot\frac{\det\Bigl[I+V\Bigr]}{\det[I+\frac1{2\pi} K]} \big| 0\rangle \, ,
\end{equation}
where the integral operator $K$ and the quantity $C(\varphi)$, which is also is expressed in
terms of certain Fredholm determinants, do not depend on $m$. The constant $C(\varphi)$
as well as the kernel $V(\lambda, \mu)$ depend on the dual fields $\varphi(\lambda)$ and $\phi(\lambda)$.
The dual fields commute for all values of spectral parameter $\lambda$. Their contribution 
to the expectation value (\ref{PnFrednonfree}) is obtained through the averaging 
procedure which suggests the decomposition of the dual fields on the relevant
creation and annihilation parts and then moving all exponentials of annihilation
operators to the right, picking up contributions whenever passing by a creation
operator. 

The general strategy of using Slavnov's formula (\ref{PnFrednonfree}) can be formulated as the following
two step procedure. The first step consists in finding the large $m$ asymptotics 
of $\det\Bigl[I + V\Bigr]$ when treating $\phi(\lambda)$ as a usual function. The second step would consist in averaging of the asymptotic formulae
obtained in the first step over the dual field vacuum{\footnote{This strategy had already been used in the two point correlation
function in the case of the 1D Bose gas at the finite coupling \cite{is,SlavnovComputationDualFieldVaccumExpLongTimeDistTempeRedDensNLSE} - another fundamental non-free fermion model .}}.
In this section of the current work we will pass through the first step.

The kernel (\ref{010}) is of the  type (\ref{definition noyau GSK}) with 
 $$
 e(\lambda) = \lambda^{\frac{m}{2}} \di e^{\frac{t \phi(\lambda)}{2}},
$$ 
and with the contour of integration being the arc $\Gamma_{\alpha}$ instead of the interval $[a;b]$. Formally, the results of
the previous sections are not directly applicable to kernel (\ref{010}). However, as we will see in  Section  \ref{maptoint}
(see Remark \ref{4.0}),
one can map this kernel to a kernel having exactly  the structure given in (\ref{definition noyau GSK}). Hence, in principle,
it is possible to  use the general  asymptotic results of Section \ref{genas}.

 At the same time, the kernel (\ref{010}) is very close to the integrable kernel
studied in \cite{dik}. Indeed, the latter is the particular case of the former corresponding $\phi(\lambda) \equiv 0$
(or, $t=0$).
Moreover, as we will see below, most of the results  and the constructions of \cite{dik}, after some minimal  modification,
can be used in the case $\phi(\lambda) \nequiv 0$ ($t>0$). This observation allows us to simplify greatly the evaluation of the
large $m$ asymptotics of  $\det[I+V]$. Basically, the only additional
analytical ingredient which is needed is the relevant differential identity for $\det[I+V]$ and some modifications
in constructing  global and local parametrices of the solution of the corresponding Riemann-Hilbert problem.
The mentioned differential identity and parametrix constructions can be extracted from the general analysis of Section \ref{genas}.

 \subsection{The $\chi$ - RH problem}
 Although not exactly of the form (\ref{definition noyau GSK}),  the kernel (\ref{010}) is still of integrable type. Therefore the arguments of Section \ref{SousSection Initial RHP} are
 applicable and we can associate with  this kernel the Riemann-Hilbert problem which consists in finding  
the $2\times 2$ matrix valued function
$\chi(\lambda)$ satisfying the following properties:
 \begin{itemize}
\item  $\chi(\lambda) \in \mathcal{O}(\C \setminus \Ga_{\alpha})$ and  has continuous  boundary values $\chi_{\pm}(\lambda)$ on $\Gamma_{\alpha}\setminus \{e^{\pm i \alpha}\} \, ;$
 \item $\chi_-(\lambda)= \chi_+(\lambda) G_{\chi}(\lambda)$ for $\lambda \in \Gamma_{\alpha}\setminus \{e^{\pm i \alpha}\}$, where 
 
\begin{equation}\label{GY}
G_{\chi}(\lambda) = I_2 +  2i\pi  \bs{E}_{R}(\la) \cdot  \bs{E}_{L}^{\bs{T}}(\la) = \begin{pmatrix}
   2 & -\lambda^{ m} \di e^{t \phi(\lambda)} \\
   \lambda^{ -m} \di e^{-t \phi(\lambda)} & 0
  \end{pmatrix} \, ; 
  \end{equation}
  
  \item $\chi(\lambda) = \log |\lambda -e^{\pm i \alpha}| \cdot  \e{O}\pa{ \ba{cc} 1 & 1 \\ 1 & 1 \ea}$ as $\lambda \rightarrow e^{\pm i \alpha}\,;$

 \item $\chi(\infty) = I_2 \,$.
\end{itemize} 
As in Section \ref{SousSection Initial RHP} (cf. (\ref{formules reconstruction chi chi-1 en terms F R et FL})), the  unique solution of this Riemann-Hilbert problem, which we will from now on  call $\chi$ - RH problem,  admits the Cauchy representations, 
  \begin{equation}\label{Ydef} 
 \chi(\lambda)= I_2 - \int_{\Ga_{\alpha}} \frac{\bs{F}_{R}(\mu)\bs{E}_{L}^{\bs{T}}(\mu)}{\mu-\lambda} \dd \mu, \qquad \mbox{and} \qquad
 \chi^{-1}(\lambda)= I_2 + \int_{\Ga_{\alpha}} \frac{\bs{E}_{R}(\mu)\bs{F}_{L}^{\bs{T}}(\mu)}{\mu-\lambda} \dd \mu \, ,
 \end{equation}
 in terms of the solutions, $\bs{F}_{R}(\la) $ and $\bs{F}_{L}(\la) $ of the  linear integral equations
$$
\bs{F}_{R}(\la) \; + \; \int_{\Ga_{\alpha}} V(\mu,\la) \bs{F}_{R}(\mu) \dd \mu \; =\;   \bs{E}_{R}(\la) \, ,
\qquad \mbox{and} \qquad 
\bs{F}_{L}(\la)  \; + \; \int_{\Ga_{\alpha}} V(\la,\mu)\bs{F}_{L}(\mu)\dd\mu  \; =  \;  \bs{E}_{L}(\la) \;. 
$$
Conversely, the vector functions  $\bs{F}_{R}(\la)$ and $\bs{F}_{L}(\la)$  are given in terms of $\chi(\lambda)$ by the equations,
 \begin{equation}\label{FHY}
\hspace{0.3cm} \bs{F}_{R}(\lambda)=\chi(\lambda)\bs{E}_{R}(\lambda)\, , \qquad \mbox{and} \qquad  \bs{F}_{L}^{\bs{T}}(\lambda)= \bs{E}_{L}^{\bs{T}}(\lambda)\chi^{-1}(\lambda)\;. 
\end{equation}

\subsection{The differential identities for the determinant}
It has already been noticed that the kernel (\ref{010}) is not exactly of the  form (\ref{definition noyau GSK}). However,
one can repeat the arguments of \cite{KozKitMailSlaTerRHPapproachtoSuperSineKernel}
 and arrive at an analogous formula to (\ref{formule derivee t determinant}) for the logarithmic  derivative of the determinant
 $\ddet{}{I + V}$. Notice that in the case of kernel (\ref{010}) we have
 $$
 p(\lambda, t)  = -i\ln \lambda -\frac{it}{m}\phi(\lambda),
 $$
 and  therefore the general identity (\ref{formule derivee t determinant}) is specified to the identity 
\begin{equation}\label{07}
\Dp{t} \ln \ddet{}{I + V} =  \frac{1}{4\pi i} \int_{\mathcal{C}} \phi(\lambda) 
\e{tr} \Big[ \Dp{\la}\chi(\la,t) \cdot \sg_3 
\cdot \chi^{-1}(\la,t) \Big]\dd\lambda \, .
\end{equation}
Here, $\mathcal{C}$ is a small counterclockwise loop around the arc $\Gamma_{\alpha}$. 

Formula (\ref{07}) reduces the asymptotic evaluation of $\ddet{}{I+V}$ to the evaluation of the uniform in $t$ asymptotics
of the solution of the $\chi$ - RH problem and the calculation of $\ddet{}{I+V}_{\mid t=0}$ which was achieved in \cite{dik}.

\subsection{The Riemann-Hilbert analysis}
The goal of this section is to produce the asymptotic solution of the $\chi$ - RH problem. This Riemann-Hilbert
problem is very close to the Riemann-Hilbert problem that was studied  in \cite{dik}. In fact,
if we put $\phi(\lambda) \equiv 0$, then $\chi(\lambda)$ will be 
 the solution of the Riemann-Hilbert problem whose asymptotics 
has been obtained in \cite{dik} (the $m$ - RH problem of \cite{dik}). It turns
out that the presence of the nontrivial  phase function $\phi(\lambda)$ does not   affect the analysis 
of \cite{dik} much, so that  we will be able to use most of the results obtained in the case
$\phi(\lambda) \equiv 0$ and to shorten  our analysis  considerably. In the rest of this section we follow the steps
used in \cite{dik} with proper technical modifications which we will handle with the help of the general analysis
of Section \ref{genas}.

\subsubsection{Mapping onto a fixed interval}\label{maptoint}

Similar to Section 3.1 of \cite{dik}, we define the linear fractional transformation $z(\la)$ by the formulae
\begin{equation}\label{9}
 	z(\la) = -i \cot\frac{\alpha}{2} \frac{\lambda-1}{\lambda + 1}\, , \qquad \mbox{viz.} \qquad  \lambda(z) = 
	\frac{1+i z\tan\frac{\alpha}{2} }{1-i z\tan\frac{\alpha}{2}}\, .
 \end{equation} 
 This change of variable transforms the $\chi-$RH problem to the following Riemann-Hilbert problem which we call the  $Y$-RH problem posed on the interval $(-1;1)$, traversed from $-1$ to $1$ :
  \begin{itemize}
 \item $Y(z) \in \mathcal{O}(\C \setminus [-1;1])\, ;$
 \item  $Y_-(z)= Y_+(z) G_{Y}(z) \, , \quad z \in (-1;1)\,$, where
 \begin{equation}\label{phyjamp}
  G_{Y}(z) = \begin{pmatrix}
    2 & -\left( \frac{1+i z \tan(\frac{\alpha}{2}) }{1-i z \tan(\frac{\alpha}{2})} \right) ^{ m} \di e^{t \phi(\lambda(z))} \\
    \left( \frac{1+i z \tan(\frac{\alpha}{2}) }{1-i z \tan(\frac{\alpha}{2})} \right) ^{ -m} \di e^{-t \phi(\lambda(z))} & 0
  \end{pmatrix}\,;
  \end{equation} 
\item $Y(\lambda) =   \log|z\mp1|  \cdot  \e{O}\pa{ \ba{cc} 1 & 1 \\ 1 & 1 \ea} $, as $z \rightarrow \pm 1\,;$  
\item $Y(\infty) = I_2\,$.
 \end{itemize}
 Once we have the solution $Y(z;m,t)$ of the $Y-$RH problem, we can find the solution $\chi(\lambda ; m,t)$ of the $\chi-$RH problem according to the  equation 
\begin{equation}\label{08}
	 \chi(\lambda ; m,t) = \left(Y(-i \cot\frac{\alpha}{2};m,t)\right)^{-1} Y(z(\lambda);m,t)
\end{equation}

\begin{rem}\label{4.0} One can notice that up to the replacement, $z \to \la$,  the $Y$ - problem is a particular case of the generalized sine kernel $\chi$ - problem  (\ref{genchiRH})
with the following specifications,
$$
p(z) = -i \ln\frac{1 + iz\tan\frac{\alpha}{2}}{1 - iz\tan\frac{\alpha}{2}}\,,\qquad \mbox{and}\qquad g(z) = t\phi(\lambda(z))\,.
$$
 \end{rem}

\subsubsection{$r$ - function transformation}

Following  \cite{dik} again, we introduce the $r$-function (this is the $g$ - function of \cite{dik} - see equation  (24) there), 
\begin{equation}\label{gfunction}
r(z) := \frac{1+i\sqrt{z^2-1}\sin(\alpha/2)}{1+iz \tan(\alpha/2)} \, .
\end{equation}
The branch of the square root is  fixed by the condition $$\sqrt{z^2-1} \sim z\,, \qquad z \to \infty \, . $$
Let us list the key properties of the  $r-$function (cf. Section 3.2 of  \cite{dik})  : 
    
    (i) \ \ \ \hspace{0.08cm}$r(z)$ is holomorphic for all $z \notin [-1;1]$.
    
    (ii) \ \ $r(z) \neq 0$ for all $z \notin [-1;1]$. At the points $z=-i \cot(\alpha/2)$ (or $z = \infty$) and $z= i \cot(\alpha/2)$ (or $z = 0$) the values of the function $r(z)$ are : 
    \begin{equation}\label{070}
    	r(-i \cot(\alpha/2)) = 1\, , \qquad \mbox{and} \qquad  r(i \cot(\alpha/2))=\cos^2(\alpha/2) \equiv \kappa\,.
    \end{equation}

(iii) \ \ The boundary values $r_{\pm}(z), z \in [-1;1]$ satisfy the following equations : 
$$ r_+(z)r_-(z) = \kappa \frac{1-i z \tan (\alpha/2)}{1+i z \tan (\alpha/2)}\, , $$ and $$\frac{r_+(z)}{r_-(z)}= \frac{1-\sqrt{1-z^2}\sin(\alpha/2)}{1+\sqrt{1-z^2}\sin(\alpha/2)}\,.$$
Here,
$$
0< \sqrt{1- z^2} \equiv   -i \lim_{\epsilon \to 0^+}\sqrt{(z +i\epsilon)^2 -1}\,, \qquad z \in (-1;1)\,,
$$
is the ``$+$"  boundary value of the function$\sqrt{z^2 -1}$ on the  segment $(-1;1)$, 
oriented from   left to  right.
This, in particular,  means that for any fixed $0 < \delta < 1 $, the following inequality holds 
\begin{equation}\label{gles1}
\left|\frac{r_+}{r_-}\right| \leq \ep_0<1, \ \ \ \ z \in [-1+\delta , 1- \delta]
\end{equation} 
for some $\ep_0=\ep_0(\delta)>0$.

(iv) \ \ The behaviour of $r(z)$ as $z \to \infty$ is described by the asymptotic relation $$ r(z) = \cos(\alpha/2) + \mathcal{O}(\frac{1}{z}) \, . $$

These properties suggest to transform the original Riemann-Hilbert problem by the formula,
  \begin{equation}\label{09}
    Y(z) \rightarrow 	\Phi(z) \equiv Y(z) r^{-m \sigma_3}(z) \kappa^{\frac{m}{2}\sigma_3} \, . 
    \end{equation} 
The matrix valued function $\Phi(z) \equiv \Phi(z;m,t)$   is the solution of the following Riemann--Hilbert problem,  which we call the $\Phi-$ RH problem :
\begin{itemize}
\item $\Phi(z) \in \mathcal{O}(\C \setminus [-1;1])\,;$ 
 \item $ \Phi_-(z)= \Phi_+(z) G_{\Phi}(z)\,, \quad z \in (-1;1)$\,, where
 \begin{equation}\label{Lajump}
G_{\Phi}(z) = \begin{pmatrix}
        2\left( r_+(z)/r_-(z) \right)^m & -e^{t \phi(\lambda(z))} \\
        e^{-t \phi(\lambda(z))} &  0
          \end{pmatrix}\,;
 \end{equation}
\item $\Phi(z) = O\Bigl( \log|z\mp1|\Bigr)$\,, \quad as $z \rightarrow \pm 1\,;$         
\item $\Phi(\infty) = I_2\,.$
\end{itemize}  
Our original problem is now reduced to the asymptotic solution of the  $\Phi-$ RH problem.

\begin{rem}\label{rem4.1} One can notice that the $\Phi-$ RH problem is a special case of the $\Xi$ - RH problem (\ref{XiRH})
considered in the main body of the paper, and that   the transition of the $Y$ - RH problem (\ref{phyjamp})  to the $\Phi$ - RH problem
(\ref{Lajump}) which we performed following \cite{dik},  is a particular case of the transition of the  general $\chi$ - RH problem (\ref{genchiRH})
to the   $\Xi$ -RH problem (\ref{XiRH}) done in Section \ref{sec1.2}. One  can  also check that the function $h(\lambda)$ associated
with the problem (\ref{phyjamp}) is related to the $r$ - function (\ref{gfunction}) by the equation,
$$
h(z) = \ln r(z) - \ln \cos\frac{\alpha}{2} \equiv  \ln \frac{1+i\sqrt{z^2-1}\sin(\alpha/2)}{1+iz \tan(\alpha/2)}  -  \ln \cos\frac{\alpha}{2}\,.
$$
\end{rem}

\subsubsection{Global parametrix}   
     
 In virtue of  estimate (\ref{gles1}), we have that, for every  $z \in (-1; 1)$,
 $$
 G_{\Phi}(z) \to \begin{pmatrix}
              0 & -e^{t \phi(\lambda(z))} \\
              e^{-t \phi(\lambda(z))} & 0 
             \end{pmatrix}\,,  
$$   
 as $m \to \infty $.  Hence, similar to the general generalised sine-kernel $\chi$ - RH problem,  one expects that $\Phi(z)$ 
 is approximated by its global parametrix,
 $M(z)$,  which  is the solution of the following RHP:
 \begin{itemize}
\item $M \in \mc{O}\big(\Cx\setminus \intff{-1}{1} \big)$ and has continuous boundary values
on $\intoo{-1}{1}\,;$
\item $M(z)=  \abs{ (z^2 -1) }^{-\f{1}{4}} \cdot  \e{O}\pa{ \ba{cc} 1 & 1 \\ 1 & 1 \ea}\,,$ 
\quad as $z \tend \pm 1\,;$ 
\item $M(z) = I_{2} \; + \; \e{O}\big(z^{-1}\big)\,, \quad$ as $z \tend \infty\,;$
\item $M_-(z) = M_+(z) \cdot G_{M}(z)$ for $z \in \intoo{-1}{1}$ where 
\beq
G_{M}(z) = \pa{ \ba{cc} 0 &  - \ex{t \phi(\lambda(z))} \\ 
						\ex{-t \phi(\lambda(z))}  &  0 \ea }  \;. 
\nonumber
\enq
\end{itemize}
This is, of course, exactly the $M$ - RH problem (\ref{MRH}) from Section \ref{glpar} with the specifications $b=-a =1$, and whose solution
is given by the equation (cf. (\ref{Msol})),
\beq\label{Msol2}
M(z) = D_{\infty}^{-\sg_3}\cdot N(z) \cdot D^{\sg_3}(z)\,,
\enq
where, given $U$ as in \eqref{definition matrices U}, 
\begin{equation}\label{Ndef}
N(z) = U^{-1} \cdot \Big( \f{z+1}{z-1}\Big)^{ \f{\sg_3}{4} } \cdot  U \,,
\end{equation}
and  (cf. (\ref{DDinfty})),
\begin{equation}\label{Ddef}
D(z) = e^{-t\eta(z)},\quad\mbox{and}\quad  D_{\infty} = e^{-t\eta_{\infty}},
\end{equation}
with 
\begin{equation}\label{etadef}
\eta(z) = -\frac{\sqrt{z^2-1}}{2 \pi } \int^{1}_{-1} \frac{\phi(\lambda(s))}{\sqrt{1- s^2} (s-z) }\dd s\,,
\end{equation} 
and
\begin{equation}\label{etainfty}
\eta_{\infty} = \frac{1}{2 \pi } \int^{1}_{-1} \frac{\phi(\lambda(z))}{\sqrt{1- z^2} }\dd z\,.
\end{equation}

The relation of the parametrix $M(z)$ to the exact solution of the $\Phi$ - RH problem is presented
in the following theorem.

 \begin{theorem}\label{th1}
    Let $\delta$ be a positive number less than $\frac{1}{4}$. Introduce the domain 
    \[
     \Om^{(\delta)} =  \Cx \setminus \ov{\mc{D}}_{1,\de} \cup \ov{\mc{D}}_{-1,\de}\,, 
     \] 
 where $\mc{D}_{1,\de}$ and $\mc{D}_{-1,\de} $ are the open balls with radius $\delta$ centered at $z=1$ and $z=-1$ respectively. 
 Then for any $t_0 \geq 0$ and  $\delta$ sufficiently small, there exist $s_0$  such that for all  $m \geq s_0$, the solution of the $\Phi$-RH problem uniquely exists and satisfies the estimate \begin{equation}\label{0}
    \Phi(z;m,\alpha)= \left( I+ O\left( \frac{1}{m(1+|z|)} \right)  \right) M(z), \ \ \ \rho \to \infty\,,
    \end{equation} uniformly for $z \in \Om^{(\delta)}$ and $ 0\leq t \leq t_0$ .     
 \end{theorem}
 The statement of this  theorem is just a particular case of the asymptotic formula (\ref{Xias}) which was proven in the main body
 of the paper for the solution of the generalised sine - kernel Riemann-Hilbert problem (\ref{genchiRH}). Alternatively, the proof can be 
 produced   by literally repeating the proof given in    \cite{dik}  for the case $\phi(\lambda) \equiv 0$. One would only need
 to use the more general local parametrices at the points $z = 1$ and $z=-1$ which in turn can be taken as special cases of the 
 parametrices constructed above  in Sections \ref{a} and \ref{b}.

\subsection{Asymptotics of the determinant}

We shall use the differential identity (\ref{07}) and Theorem \ref{th1}. Technically, this is simpler than to try to  extract the
result from our general  formula (\ref{detgen}) of Proposition \ref{prop1}.  Tracing back the chain of transformations that led us from
the original function $\chi(\lambda)$ to the function $\Phi(z)$ we have that
\begin{equation}\label{chiPhi}
\chi(\lambda) =\kappa^{\frac{m}{2}\sigma_3}\Phi^{-1}\Bigl(-i\cot\frac{\alpha}{2}\Bigr)\Phi(z(\lambda))
r^{m\sigma_3}(z(\lambda))\kappa^{-\frac{m}{2}\sigma_3}.
\end{equation}
This would yield the following expression for the product $\chi^{-1}(\lambda)\partial_\lambda\chi(\lambda)$
which is involved in the integral in the right hand side of (\ref{07}),
\beq
\chi^{-1}(\lambda)\partial_\lambda\chi(\lambda)=m\partial_\lambda r(z(\lambda)) r^{-1}(z(\lambda))\sigma_3   
+\kappa^{\frac{m}{2}\sigma_3}r^{-m\sigma_3}(z(\lambda))\Phi^{-1}(z(\lambda))
\partial_\lambda\Phi(z(\lambda))r^{m\sigma_3}(z(\lambda))\kappa^{-\frac{m}{2}\sigma_3},
\nonumber
\enq
and, in turn,
\begin{equation}\label{asdet0}
\e{tr} \Big[ \Dp{\la}\chi(\la,t) \cdot \sg_3 
\cdot \chi^{-1}(\la,t) \Big]
 =  2m\partial_\lambda r(z(\lambda))r^{-1}(z(\lambda))
+  \e{tr} \Big[ \sigma_3\Phi^{-1}(z(\lambda))
\partial_\lambda\Phi(z(\lambda))\Big]\,.
\end{equation}

As it follows from Theorem \ref{th1}, on  the loop $\mathcal{C}$ the function  $\Phi(z(\lambda))$ can be approximated by the the global parametrix
$M(z(\lambda))$. Indeed, from (\ref{0}) we have that
\begin{equation}\label{asdet1}
\Phi(z(\lambda)) = \left(I_2 + O\left(\frac{1}{m}\right)\right)M(z(\lambda)),\quad m \to \infty,
\end{equation}
where the estimate holds uniformly for $\lambda \in \mathcal{C}$, and are  differentiable with respect to $\lambda$. 
Combining (\ref{asdet1}) and (\ref{asdet0}) we conclude that
\begin{equation}\label{asdet2}
\e{tr} \Big[ \Dp{\la}\chi(\la,t) \cdot \sg_3 
\cdot \chi^{-1}(\la,t) \Big]
 =  2m\partial_\lambda r(z(\lambda))r^{-1}(z(\lambda))
+  \e{tr} \Big[ \sigma_3M^{-1}(z(\lambda))
\partial_\lambda M(z(\lambda))\Big]
 + O\left(\frac{1}{m}\right).
\end{equation}
Using the definition (\ref{Msol2}) of the global parametrix $M(z(\lambda))$, we derive from (\ref{asdet2}) the
following  asymptotic formula for  the  integrand (\ref{07}) expressed in terms of  the known objects,
\beq
\e{tr} \Big[ \Dp{\la}\chi(\la,t) \cdot \sg_3 
\cdot \chi^{-1}(\la,t) \Big]
 =  2m\partial_\lambda r(z(\lambda))r^{-1}(z(\lambda)) - 2t\partial_{\lambda}\eta(z(\lambda))
+  \e{tr} \Big[ \sigma_3N^{-1}(z(\lambda))
\partial_\lambda N(z(\lambda))\Big]
 + O\left(\frac{1}{m}\right),
\label{asdet3}
 \end{equation}
 where the functions $\eta(z)$ and $N(z)$ are given by the equations (\ref{etadef}) and (\ref{Ndef}),
 respectively. Observe that
 \begin{equation}\label{asdet4}
 \e{tr} \Big[ \sigma_3N^{-1}(z(\lambda))
\partial_\lambda N(z(\lambda))\Big]
 =  \frac{1}{4}\partial_\lambda\ln\beta(\lambda) \e{tr} \Big[ \sigma_3 U^{-1}\sigma_3 U\Big] = 0\,,
\end{equation}
where
$$
\beta(\lambda) = \frac{z(\lambda) + 1}{z(\lambda) - 1} \, . 
 $$
 Therefore, the asymptotic 
formulae (\ref{asdet3}) reduces to the relation
\begin{equation}\label{asdet5}
\e{tr} \Big[ \Dp{\la}\chi(\la,t) \cdot \sg_3 
\cdot \chi^{-1}(\la,t) \Big]
=2m\partial_\lambda r(z(\lambda))r^{-1}(z(\lambda)) - 2t\partial_{\lambda}\eta(z(\lambda))     
 + O\left(\frac{1}{m}\right),
 \end{equation}
as $m \to \infty$,  uniformly for  $\lambda \in \mathcal{C}$.

Substituting the estimate  (\ref{asdet5}) into the right hand side of (\ref{07}) and changing the variable
of integration, $\lambda\to z$,  we obtain that
$$
\Dp{t} \ln \ddet{}{I + V}=  \frac{m}{2\pi i} \int_{\mathcal{L}} \phi(\lambda(z)) \partial_{z}\ln r(z)\dd z
-\frac{t}{2\pi i} \int_{\mathcal{L}} \phi(\lambda(z)) \partial_{z}\eta(z)\dd z
+ O\left(\frac{1}{m}\right),\quad m \to \infty\,,
$$ 
where $\mathcal{L}$ is a small loop around the interval $[-1, 1]$ and the estimate is uniform with respect to $t$. 
Integrating this estimate, we arrive at the following  asymptotics for the  determinant,
\begin{equation}\label{asdet6}
\ln \ddet{}{I + V} = \ln \ddet{}{I + V}\Bigl|_{t=0}  +  \frac{mt}{2\pi i} \int_{\mathcal{L}} \phi(\lambda(z)) \partial_{z}\ln r(z)\dd z
-\frac{t^2}{4\pi i} \int_{\mathcal{L}} \phi(\lambda(z)) \partial_{z}\eta(z)\dd z
+ O\left(\frac{1}{m}\right),\quad m \to \infty\,,
\end{equation}
Using the known \cite{w} (see also \cite{dik}) large $m$  asymptotics of the $\ln \ddet{}{I + V}\Bigl|_{t=0} $, we transform (\ref{asdet6})
into our final asymptotic result,
\bem
\ln \ddet{}{I + V}  = m^2\ln\cos{\frac{\alpha}{2}} + 
 m\frac{t}{2\pi i} \int_{\mathcal{L}} \phi(\lambda(z)) \partial_{z}\ln r(z)\dd z    \\ 
 - \frac{1}{4} \ln \left(m\sin{\frac{\alpha}{2}}\right)
-\frac{t^2}{4\pi i} \int_{\mathcal{L}} \phi(\lambda(z)) \partial_{z}\eta(z)\dd z + c_0
+ O\left(\frac{1}{m}\right),\quad m \to \infty\,,
\label{asdet7}
\end{multline}
where $c_0$ is the famous Widom's constant:
$$
c_0 = \frac{1}{12}\ln 2 + 3\zeta'(-1).
$$

\section*{Acknowledgements}
R. G. is supported by the  NSF Grant DMS-1700261.  A. R. I. is supported by the  NSF Grant DMS-1700261 and Russian Science Foundation grant No.17-11-01126..
K. K. K. is supported by CNRS. This work has been supported by the grant PEPS-PTI "Asymptotique d'int\'{e}grales multiples". 
K. K. K. would like to thank the mathematics department of IUPUI for its warm hospitality and financial support during 
his visit there when this work has been carried out.



\begin{thebibliography}{100} 
	

\bibitem{BealsCoifmanScatteringInFirstOrderSystemsEquivalenceRHPSingIntEqnMention}
R.~Beals and R.R.~Coifman, \emph{{"Scattering and inverse scattering for first order systems."}}, Comm. Pure Appl. Math. \textbf{\bf{37}:1}
  (1984), 39-90.


  
  
  \bibitem{CalderonContinuityCauchyTransformLipschitzCurves}
A.P.~Calderon, \emph{{"Cauchy integrals on Lipschitz curves and related operators."}}, oProc. Natl. Acad. Sci. USA \textbf{\bf{74}:4}
  (1977), 1324-1327.


\bibitem{DeiftItsZhouSineKernelOnUnionOfIntervals}
P.A.~Deift and A.R.~Its and X.~Zhou, \emph{{"A Riemann-Hilbert approach to asymptotics problems arising in the theory of random matrix models
    and also in the theory of integrable statistical mechanics."}}, Ann. Math. \textbf{\bf{146}}
  (1997), 149--235.

\bibitem{DeiftZhouSteepestDescentForOscillatoryRHP}
P.A. Deift and X.~Zhou, \emph{{"A steepest descent method for oscillatory
  Riemann-Hilbert problems."}}, Bull. Amer. Math. Soc. \textbf{\bf{26}:1}
  (1992), 119--123.

 \bibitem{deift} P.~A. Deift, Integrable operators, in
\emph{{"Differential operators and spectral theory: M. Sh. Birman's
70th anniversary collection}}, V.~Buslaev, M.~Solomyak, D.~Yafaev, eds.,
American mathematical Society Translations, ser.~2, v.~189, Providence,
RI: AMS, 1999.

\bibitem{dik}  Deift, P., Its, A., Krasovsky, I., and Zhou, X. \emph{{"The Widom-Dyson constant for the gap probability in
	random matrix theory}}, JCAM, {\bf{202}}, (2007), 26-47. 

\bibitem{EhrhardtConstantinPureSinekernelFredholmDeyt}
T. Ehrhardt, \emph{{"Dyson's constant in the asymptotics of the Fredholm determinant of the sine kernel."}}, Comm. Math. Phys.
  \textbf{\bf{262}} (2006), 317-341.

  
 
\bibitem{ItsIzerginKorepinSlavnovDifferentialeqnsforCorrelationfunctions}
A.R. Its, A.G. Izergin, V.E. Korepin, and N.A. Slavnov, \emph{{"Differential
  equations for quantum correlation functions."}}, Int. J. Mod. Physics
  \textbf{\bf{B4}} (1990), 1003--1037.

\bibitem{is} A. R. Its and N. A. Slavnov, \emph{{"On the Riemann-Hilbert approach to the asymptotic analysis
of the correlation functions of the Quantum Nonlinear Schrodinger equations. Non-free fermonic case.}}, Theor. Math. Phys. {\bf{119}:2}, (1999), 541-593



\bibitem{itw} A. Its, C. Tracy, H. Widom,  \emph{{"Random Words, Toeplitz Determinants and
   Integrable Systems. II}}, {\it  Physica D} 152-153 (2001), 199-224.


    \bibitem{KitanineMailletTerrasElementaryBlocksPeriodicXXZ}
N.~Kitanine and J.-M.~Maillet and V.~Terras, 
  \emph{{"Correlation functions of the XXZ Heisenberg spin-$1/2$ chain in a magnetic field."}}, Nucl. Phys. B.  \textbf{\bf 567} , (2000),
  554--582.	

\bibitem{KozKitMailSlaTerXXZsgZsgZAsymptotics}
N.~Kitanine, K.K. Kozlowski, J.-M. Maillet, N.A. Slavnov, and V.~Terras,
  \emph{{"Algebraic Bethe Ansatz approach to the asymptotics behavior of
  correlation functions."}}, J. Stat. Mech: Th. and Exp. \textbf{04} (2009),
  P04003.

\bibitem{KozKitMailSlaTerRHPapproachtoSuperSineKernel}
N.~Kitanine, K.K. Kozlowski, J.-M. Maillet, N.A. Slavnov, and V.~Terras, 
\emph{{"The Riemann-Hilbert approach to a generalized sine kernel and
  applications."}}, Comm. Math. Phys. \textbf{{\bf 291} :3} (2009), 691--761.

\bibitem{KozReducedDensityMatrixAsymptNLSE}
K.K. Kozlowski, \emph{{"Large-distance and long-time asymptotic behavior of the
  reduced denisty matrix in the non-linear Schr\"{o}dinger model."}}, to appear
  in Ann. Henri-Poincar\'{e} \textbf{16,3} (2015).
  
\bibitem{KozProofOfDensityOfBetheRoots}
K.K.~Kozlowski, \emph{{"On condensation properties of Bethe roots associated with the XXZ chain."}}, Comm. Math. Phys.  \textbf{\bf 357} ,3, (2018),
  1009-1069.	
    
\bibitem{KozTerNatteSeriesNLSECurrentCurrent}
K.K. Kozlowski and V.~Terras, \emph{{"Long-time and large-distance asymptotic
  behavior of the current-current correlators in the non-linear Schr\"{o}dinger
  model."}}, J. Stat. Mech.: Th. and Exp. (2011), P09013.

\bibitem{kras} I. V. Krasovsky, Gap probability in the spectrum of random matrices and saymptotics of
polynomials orthogonal on an arc of the unit circle, {\it Int. Math. Res. Not.} {\bf 2004} (2004), 1249 - 1272.


\bibitem{w} H. Widom, \emph{{"The strong Szeg\H{o} limit theorem for circular arcs,}} {\it Indiana UNiv. Math. J.} {\bf 21} (1971), 277 - 283.
 
    
\bibitem{KuilajaarsMVVUniformAsymptoticsForModifiedJacobiOrthogonalPolynomials}
A.B.J. Kuijlaars, K.T.-R. McLaughlin, W.~Van Assche, and M.~Vanlessen,
  \emph{{"The Riemann-Hilbert approach to strong asymptotics for orthogonal
  polynomials on [-1,1]."}}, Advances in Math. \textbf{\bf 188} (2004),
  337--398.	
	

\bibitem{SlavnovComputationDualFieldVaccumExpLongTimeDistTempeRedDensNLSE}
N.A.~Slavnov
  \emph{{"Integral equations for the correlation functions of the quantum one-dimensional Bose gas."}}, Theor. Math. Phys. \textbf{\bf 121} (1999),
  1358--1376.
	
	
	  

  
\end{thebibliography}

\appendix

\section{Special functions}
\label{Appendix}

We gather in the appendix some elementary information about Hankel functions. The Hankel functions satisfy to the addition formulae
\beq
H_0^{(1)}(\ex{-i\pi} z)  = 2 H_0^{(1)}(z) \; + \; H_0^{(2)}(z) \qquad  \e{and}  \qquad 
H_0^{(2)}(\ex{-i\pi}z) = - H_0^{(1)}(z) \;, 
\label{Appendix ecriture saut Hankel 1}
\enq
from which follow formulae for additions of derivatives
\beq
\big[ H_0^{(1)} \big]^{\prime} (\ex{-i\pi} z)  = - 2 \big[H_0^{(1)}\big]^{\prime}(z) \; - \; \big[H_0^{(2)}\big]^{\prime}(z) 
\qquad  \e{and}  \qquad 
\big[H_0^{(2)}\big]^{\prime}(\ex{-i\pi}z) =  \big[ H_0^{(1)} \big]^{\prime} (z) \;.
\label{Appendix ecriture saut Hankel 1 prime}
\enq
Similar results follow for the rotations by $\ex{i\pi}$. 
\beqa
H_0^{(2)}(\ex{i\pi} z)  = 2 H_0^{(2)}(z) \; + \; H_0^{(1)}(z) \quad & \e{and} & \quad 
H_0^{(1)}(\ex{i\pi}z) = - H_0^{(2)}(z) \;, 
\label{Appendix ecriture saut Hankel 2}\\
\big[ H_0^{(2)} \big]^{\prime} (\ex{i\pi} z)  = - 2 \big[H_0^{(2)}\big]^{\prime}(z) \; - \; \big[H_0^{(1)}\big]^{\prime}(z) 
\quad & \e{and}  & \quad 
\big[H_0^{(1)}\big]^{\prime}(\ex{i\pi}z) =  \big[ H_0^{(2)} \big]^{\prime} (z) \;.
\label{Appendix ecriture saut Hankel 2 prime}
\eeqa
These function exhibit the local behavior at $z=0$: $H_0^{(a)}(z)=\e{O}(\ln z) $ with $a=1,2$.

The Hankel functions admit the $z \tend \infty$ asymptotic expansions
\beqa
H_{\nu}^{(1)}(z) &\simeq & -i \Big( \f{2i}{\pi z} \Big)^{\f{1}{2}} \ex{iz} \ex{-i\f{\pi \nu}{2} }
\sul{n=0}{+\infty} \Big( \f{i}{ 2 z} \Big)^{n} (\nu,n)   \;\; ,  \label{Appendix Asympt Exp Hankel 1}\\
H_{\nu}^{(2)}(z) &\simeq & i \Big( \f{- 2i}{\pi z} \Big)^{\f{1}{2}} \ex{-iz} \ex{i\f{\pi \nu}{2} }
\sul{n=0}{+\infty} \Big( \f{ -i }{ 2 z} \Big)^{n} (\nu,n)  \;\; , \label{Appendix Asympt Exp Hankel 2}
\eeqa
where we agree upon 
\beq
(\nu,m) = \Ga\pab{ \nu + \tf{1}{2} + m }{  \nu + \tf{1}{2} - m, m+1 } \;. 
\label{definition symbole nu n}
\enq

These asymptotic expansions  being differentiable, we infer that
\beqa
\big[ H_{0}^{(1)} \big]^{\prime}(z) &\simeq &  \Big( \f{2i}{\pi z} \Big)^{\f{1}{2}} \ex{iz} 
\sul{n=0}{+\infty} \Big( \f{i}{ 2 z} \Big)^{n} (0,n) \f{1+2n}{1-2n}  \label{Appendix Asympt Exp Hankel prime 1}\\
\big[ H_{0}^{(2)} \big]^{\prime}(z) &\simeq & \Big( \f{-2i}{\pi z} \Big)^{\f{1}{2}} \ex{-iz} 
\sul{n=0}{+\infty} \Big( \f{ -i }{ 2 z} \Big)^{n} (0,n) \f{1+2n}{1-2n} \;. 
\label{Appendix Asympt Exp Hankel prime 2}
\eeqa
%
%

\bibliographystyle{Plain}

\end{document}